\documentclass[twocolumn,trackchanges]{aastex62}

\received{XX}
\revised{YY}
\accepted{ZZ}

\submitjournal{ApJ}
\shorttitle{The eclipsing binary DD\,CMa}
\shortauthors{Rosales, J. A.}

\usepackage{txfonts}
\newcommand{\kms}{\,km\,s$^{-1}$}

\begin{document}

\title{Fundamental parameters of the eclipsing binary DD\,CMa and evidence for mass exchange}

\correspondingauthor{Rosales, J. A.}
\email{jrosales@astro-udec.cl, jrosales@mao.kiev.ua}

\author[0000-0001-6657-9121]{Rosales, J. A.}
\affiliation{Departamento de Astronom\'{i}a, Universidad de Concepci\'{o}n, Casilla 160-C, Concepci\'{o}n, Chile.}
\affiliation{Main Astronomical Observatory, National Academy of Sciences of Ukraine, 27 Akademika Zabolotnoho St, 03680 Kyiv, Ukraine.}

\author{Mennickent, R. E.}
\affiliation{Departamento de Astronom\'{i}a, Universidad de Concepci\'{o}n, Casilla 160-C, Concepci\'{o}n, Chile.}

\author{Djura\v{s}evi\'{c}, G.}
\affiliation{Astronomical Observatory, Volgina 7, 11060 Belgrade 38, Serbia.}
\affiliation{Isaac Newton Institute of Chile, Yugoslavia Branch, 11060 Belgrade, Serbia.}

\author{Gonz\'{a}lez, J. F.}
\affiliation{Instituto de Ciencias Astron\'{o}micas, de la Tierra y del Espacio (CONICET-Universidad Nacional de San Juan), CC 49, 5400 San Juan, Argentina.}

\author{Araya, I.}
\affiliation{Centro de Investigaci\'{o}n DAiTA Lab, Facultad de Estudios Interdisciplinarios, Universidad Mayor, Chile.}

\author{Cabezas, M.}
\affiliation{Astronomical Institute of the Czech Academy of Sciences, Bo\v{c}n\'i II 1401/1, Prague, 141 00, Czech Republic.}
\affiliation{Institute of Theoretical Physics, Faculty of Mathematics and Physics, Charles University, V Holesovickach 2, 180 00 Praha 8, Czech Republic.}

\author{Schleicher, D.\,R.\,G.}
\affiliation{Departamento de Astronom\'{i}a, Universidad de Concepci\'{o}n, Casilla 160-C, Concepci\'{o}n, Chile.}

\author{Cur\'{e}, M.}
\affiliation{Instituto de F\'{i}sica y Astronom\'{i}a, Facultad de Ciencias, Universidad de Valpara\'{i}so, Chile.}

\begin{abstract}

We present a detailed photometric and spectroscopic analysis of DD\,CMa, based on published survey photometry and new spectroscopic data. We find an improved orbital period of $P_\mathrm{o}$= 2\fd0084530(6). Our spectra reveal H$\beta$ and H$\alpha$ absorptions with weak emission shoulders and we also find color excess in the WISE multiband photometry, interpreted as signatures of circumstellar matter. We model the  $V$-band orbital light curve derived from the ASAS and ASAS-SN surveys, assuming a semidetached configuration and using the mass ratio and temperature of the hotter star derived from our spectroscopic analysis.  Our model indicates that the system consists of a B\,2.5 dwarf and a B\,9 giant of radii 3.2 and 3.7 $\mathrm{R_{\odot}}$, respectively, orbiting in a circular orbit of radius 6.75  $\mathrm{R_{\odot}}$. We also found  $M_{\mathrm{c}} = 1.7 \pm 0.1 ~\mathrm{M_{\odot}}$, $T_{\mathrm{c}} = 11350 \pm 100 ~\mathrm{K}$ and $M_{\mathrm{h}} = 6.4 \pm 0.1 ~\mathrm{M_{\odot}}$, $T_{\mathrm{h}} = 20000 \pm 500 ~\mathrm{K}$, for the cooler and hotter star, respectively. We find broad single emission peaks  in  H$\alpha$ and H$\beta$ after subtracting the  synthetic stellar spectra. Our results are consistent with mass exchange between the stars, and suggest the existence of a stream of gas being accreted  onto the early B-type star.
\end{abstract}

\keywords{stars: binaries: eclipsing--binaries: close -- binaries: spectroscopic -- stars: activity-- stars: circumstellar matter-- stars: fundamental parameters}


\section{Introduction}
\label{Sec: Sec. 1}
\noindent

The first spectroscopic binary in being discovered was one of the stellar components of Mizar \citep[$\zeta$ Ursae Majoris,][]{1890MNRAS..50..296P}. These spectroscopic binaries, defined as those showing two sets of lines moving in anti-phase in their spectra,  have high astrophysical value,  since the use of well established methods including the analysis of the eclipse depth and its duration, allow us to derive the whole set of orbital and stellar parameters. Some of the current studies of the binary star populations show that the observed distributions of binary separation extends from 10$^{-2}$ to 10$^5$ AU. The explanation for this wide range of separation must lie in the details of the star formation process  \citep{2005A&A...437..113H}.  Among this huge variety of binaries, there is a group whose separation is close enough to allow interaction, through stellar wind accretion or Roche-lobe overflow \citep{2006epbm.book.....E}.  In close binary systems, mass transfer is always accompanied by an exchange of angular momentum. In cases where the transferred mass directly impacts on the accretor, the excess kinetic energy of the gas stream is radiated in an optically thick hot spot or in a hot equatorial band \citep{1976ApJ...206..509U}. In some interacting binaries,  the existence of additional light sources as accretion disk, gas stream and shock regions
complicates the task of obtaining stellar parameters from the light curve analysis only. Combined spectroscopic and photometric information is needed to constrain the physical scenarios and help to understand these stellar systems. On the other hand, the evolution of close binaries of intermediate mass is  not yet well understood, lacking the knowledge of details in the process of systemic mass loss and stellar mass exchange throughout the interaction stage \citep{2012Sci...337..444S, 2014ApJ...782....7D}. In order to have a global picture of the binary evolution, understanding these evolutionary stages is essential, since it has been shown that binary interaction dominates the evolution of massive stars and  more than 70\% of all massive stars will exchange mass with a companion during its lifetimes  \citep{2012Sci...337..444S}.

In this paper we present a study of DD\,CMa, an interesting short orbital period eclipsing binary showing signatures of mass exchange, which could be in an advanced evolutionary stage. DD\,CMa was discovered as a variable star of the Algol type after  inspecting photographic plates in the Laboratoire d` Astronomie et de G\'eod\'esie de l' Universit\'e de Louvain   \citep{1949PLAGL..12E..17D}. According to SIMBAD\footnote{\url{http://simbad.u-strasbg.fr/simbad/}}, this object is an Eclipsing Algol Semi-Detached (EA/SD) binary; it is also named  
ASAS ID 072409-1910.8, and is characterized by $\alpha_{2000}$= 07:24:09, $\delta_{2000}$=-19:10:48,  $V= 11.56 \pm 0.11$ mag and $B-V= 0.1 \pm 0.19$ mag.
The orbital period has been successively reported as $P_\mathrm{o}$= 2\fd0083807 $\pm$ 0\fd0000027, $P_\mathrm{o}$= 2\fd0083, 2\fd008452 and 2\fd0084 $\pm$ 0\fd0001 by  \citet{1949PLAGL..12E..17D}, ASAS \citep{1997AcA....47..467P}\footnote{\url{http://www.astrouw.edu.pl/asas/?page=acvs}}, the International Variable Star Index (VSX)\footnote{\url{https://www.aavso.org/vsx/}} and \citet{2017IBVS.6207....1R}. 
The distance based on the GAIA\footnote{\url{http://gea.esac.esa.int/archive/}} DR2 parallax is 2632 [+309 -251] pc \citep{2018AJ....156...58B}. The object has not been studied in detail until now. In this study we expect to determine for the first time the fundamental stellar and orbital parameters of this system and to contribute to the knowledge  of its evolutionary stage.

In Section \ref{Sec: Sec. 2} we present a new photometric analysis of DD\,CMa. In Section \ref{Sec: Sec. 3} we present new spectroscopic data acquired by us along with our methods of data reduction. In the same section we calculate the orbital parameters of the system and the physical parameters of the brighter star. In Section \ref{Sec: Sec. 4} we model the $V$-band light curve of the binary and derive a complete set of parameters including the system inclination, the  stellar separation and the absolute magnitudes, temperatures, masses and radii for both stars. In Section \ref{Sec: Sec. 5} we fit the spectral energy distribution obtaining the distance to the system. In Section \ref{Sec: Sec. 6} we discuss the H$\alpha$ residual emission. 
In Section  \ref{Sec: Sec. 7} we provide a discussion of  our results. Finally, in Section \ref{Sec: Sec. 8} we summarize the main results of our investigation.

\section{Photometric analysis}
\label{Sec: Sec. 2}

\begin{figure}
	\begin{center}
		\includegraphics[trim=0.2cm 0.0cm 0.0cm 0.0cm,clip,width=0.5\textwidth,angle=0]{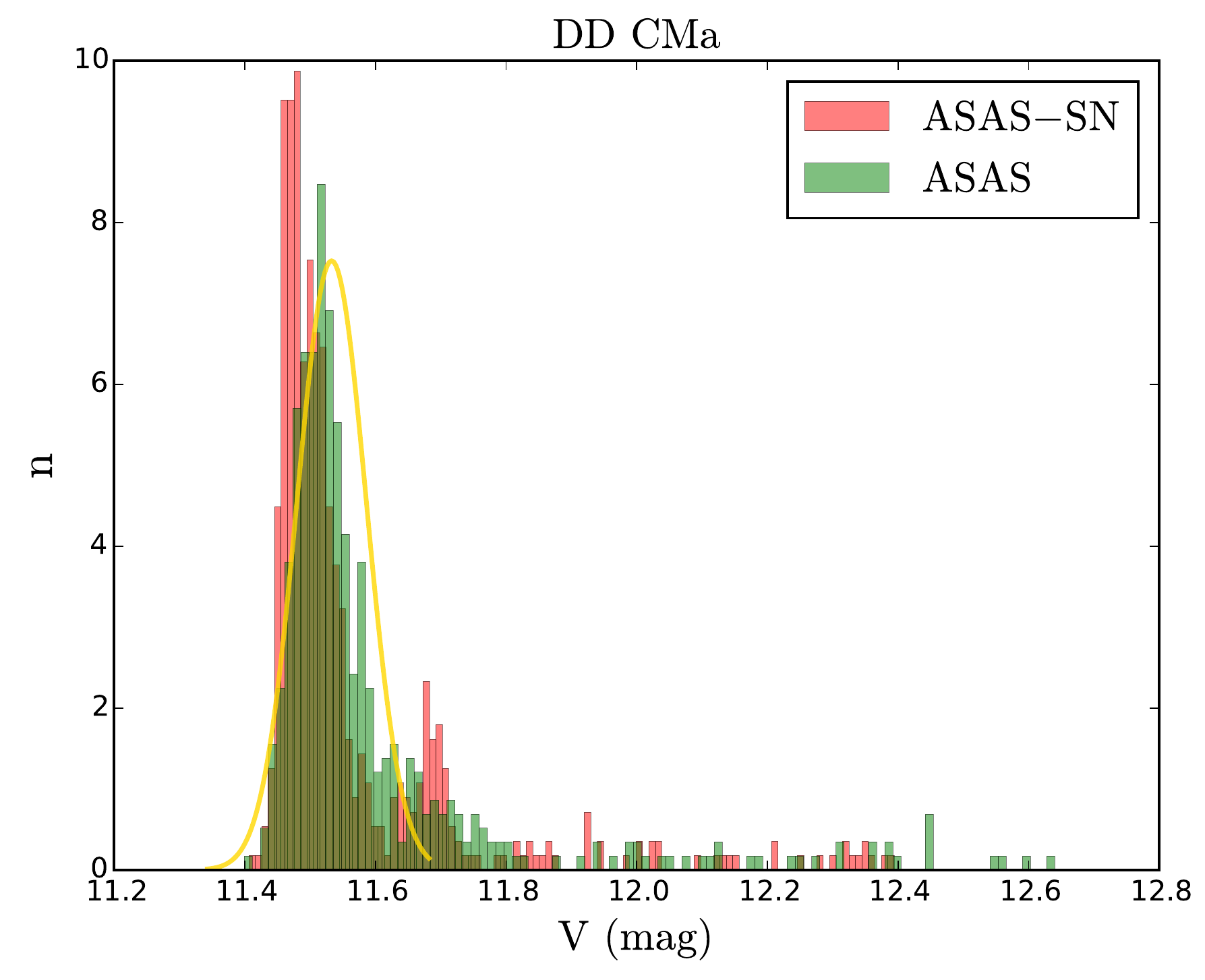}
	\end{center}
	\caption{Histograms for 466 ASAS (green) and 564 ASAS-SN (red) magnitudes. The continue yellow line shows the fit to the ASAS distribution, and corresponds to a normal distribution with mean $\mu = 11.533$ mag and sigma $\sigma = 0.053$ mag.}		
	\label{fig:Fig. 1}
\end{figure}

\noindent
In this section we present the analysis of ASAS and ASAS-SN\footnote{\url{https://asas-sn.osu.edu/variables}} \citep{2014ApJ...788...48S,2017PASP..129j4502K,2019MNRAS.486.1907J}  photometric databases. For our analysis, we have considered only the best-quality data, excluding lower quality magnitudes available in the database. These photometric time series present a skewed distribution with a tail towards faint magnitudes  (Fig. \ref{fig:Fig. 1}). Those observed skewed data in both distributions basically correspond to the magnitudes measured during the primary and secondary eclipse that range from approximately 11.6 to 12.6 mag. Then if the primary and secondary eclipses are less deep, the skews would be much smaller. We also present the analysis of eclipse timings and survey infrared photometry.

\subsection{A search for periodicities}
\label{subsec: 2.1}

The first photometric inspection based on ASAS data  was performed by \citet{2017IBVS.6207....1R}, and revealed that DD\,CMa is a binary  with a short orbital period of 2\fd00844. In addition, the above study suggested the presence of a long photometric periodicity, beyond the orbital modulation. In this subsection we re-analyze the ASAS dataset and shows that the previously reported long-cycle is an artifact produced by data sampling and orbital frequency.

With the objective of searching for  a possible additional long photometric period, we used an algorithm written by Zbigniew Kolaczkowski to disentangle the light curve. This piece of code adjusts the data with a Fourier series consisting of fundamental frequencies plus their harmonics, which are previously determined e.g. with the Phase Minimization Dispersion (\texttt{PDM}) code \citep{1978ApJ...224..953S}. The number of harmonics will depend on the form of the light curve. Then, if the light curve has a sinusoidal form could be enough to use a few number of harmonics. Since the light curve of an eclipsing binary is more complex than a sinusoidal form, it is necessary a greater amount of harmonics, typically 12 or more to fit it adequately. After adjusting the light curve with the main frequency and theirs harmonics, the code provides the residuals, that can be examined searching for additional frequencies. This code is described in more detail by \citet{2012MNRAS.421..862M}.

Following the above procedure with the light curve composed by ASAS and ASAS-SN magnitudes (466 and 564 $V$-band magnitudes, respectively), 
and applying the Generalized Lomb-Scargle (GLS) algorithm, we get an orbital period of 2\fd00844 $\pm$ 0\fd00002, confirming the result given by  \citet{2017IBVS.6207....1R}. 
In addition, we find that the residuals does not show any additional frequency.  
As mentioned before, the long-cycle previously reported is an artifact of the data sampling and orbital frequency; the overall light curve seems to follow a 89 d periodicity just because of the combination of the 2 d cycle and the night-to-night observational gaps.

\subsection{Analysis of the main-eclipse timing}
\label{subsec: 2.2}

\noindent
In order to improve the accuracy of the orbital period, we selected the photometric data points with phases close to the primary minimum 
and performed an analysis of observed (O) minus calculated (C) eclipse times using as test period  2\fd00844 and  following  \citet{2005ASPC..335....3S}.  In this analysis the O-C deviations can be represented as a function of the number of cycles with a straight line whose
zero point and slope are the corrections needed for the (linear) ephemeris zero point and test period, respectively. We include 6 times of primary minimum 
studied by \citet{1949PLAGL..12E..17D} and 16 new times of minima measured by us from the ASAS and ASAS-SN databases. 
The dataset of minima covers 15289 cycles, i.e. 
84 years and is presented in Tab.\,\ref{Tab: Tab. 1}. Since some minima times are published without errors, we use a simple least square fit for our analysis.
Our result displayed in Fig. \ref{fig:Fig. 3} shows that the fit can be performed with a straight line, indicating a constant period. 
The new ephemeris is given by: 

\begin{eqnarray}
\rm HJD_{min} = 2427537.311(7) + E\times 2\fd0084530(6)
\label{eq: eq. 1}
\end{eqnarray}

Due to the much longer time baseline considered in this paper, we find that this period is more reliable than $P_{\rm o}$= 2\fd0083807 $\pm$ 0\fd0000027, 
provided by \citet{1949PLAGL..12E..17D}. Actually, the same period given by Eq.\,\ref{eq: eq. 1} is found when doing the O-C analysis using as test 
the Deurinck's period. The ephemeris given in  Eq.\,\ref{eq: eq. 1}  will be used for the photometric and spectroscopic analysis in the rest of the paper. 

\begin{figure}
	\begin{center}
		\includegraphics[trim=0.2cm 0.4cm 0.2cm 0.2cm,clip,width=0.45\textwidth,angle=0]{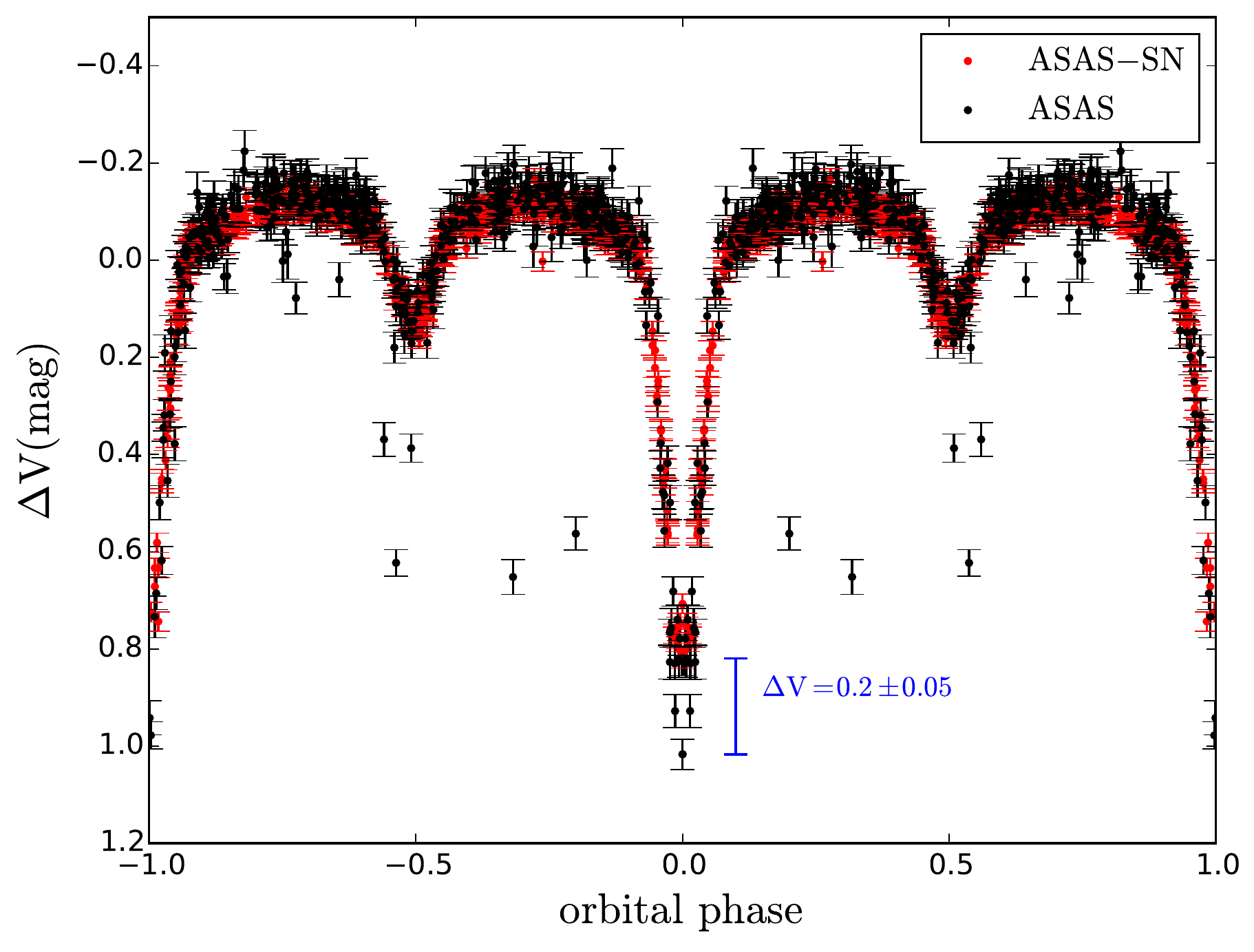}
	\end{center}
	\caption{ASAS and ASAS-SN $V$-band light curves phased with ephemerides given by Eq.\,1}    
	\label{fig:Fig. 2}
\end{figure}

\begin{figure}
	\begin{center}
		\includegraphics[width=1.0\linewidth]{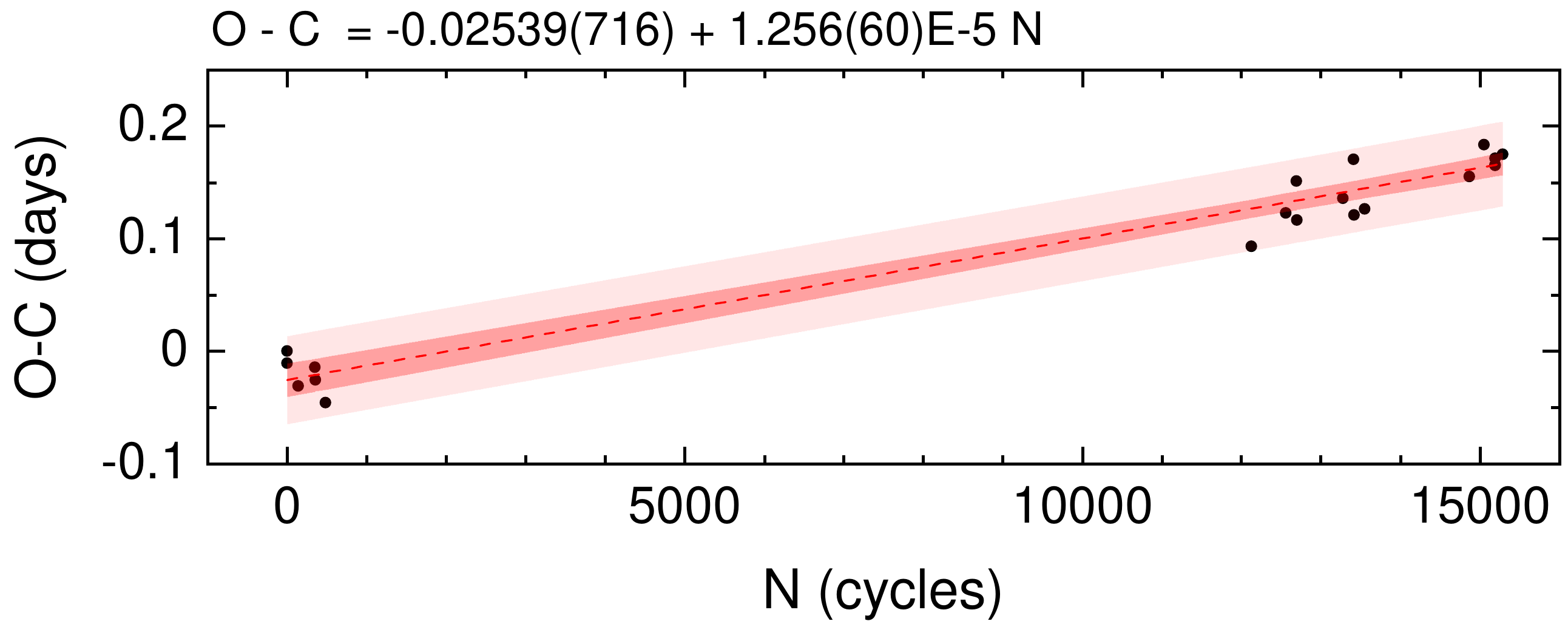}
	\end{center}
	\caption{Observed (O) minus calculated (C) epochs for primary minima versus cycle number for 84 years of observations, calculated using a test period of $P_{\rm o}$= 2\fd00844, along with the best straight line fit. The 
		dark pink region indicates the 95\% confidence level, and the light pink region the 95\% prediction level.} 
	\label{ocfit}
	\label{fig:Fig. 3}
\end{figure}

The light curve phased with the ephemerides given in Eq. \ref{eq: eq. 1} is shown in Fig.\,\ref{fig:Fig. 2}. Three points suggest a deeper main eclipse during the epochs covered by ASAS, but the larger scatter shown by ASAS data places a doubt on the significance of this finding. Actually, in other epochs we also observe - few - fainter than average ASAS magnitudes, around the secondary eclipse and around phases 0.2 and 0.35, for instance.

\begin{table}
	\caption{Times of main eclipse minima studied in this paper. Data from \citet{1949PLAGL..12E..17D} are in JD and others in HJD. }
	\label{Tab: times}
	\centering
	\begin{tabular}{l r l r }
		\hline\hline
		JD/HJD & Source & JD/HJD & Source\\
		\hline  
		2427537.336	&DEURINCK& 2454470.68677	 &ASAS       \\
		2427539.334	&DEURINCK&2454482.68797	 &ASAS        \\
		2427810.453	&DEURINCK&2454757.84982	 &ASAS         \\
		2428240.276	&DEURINCK&2457392.95175	 &ASAS-SN \\
		2428246.29	&DEURINCK&2457774.58377	 &ASAS-SN\\
		2428507.367	&DEURINCK&2457774.58384	 &ASAS-SN \\
		2451889.76435	& ASAS	    & 2458053.74451	 &ASAS-SN\\
		2452763.46515	& ASAS	    & 2458053.74487	 &ASAS-SN \\
		2453030.61594	& ASAS    	    & 2458055.74683	 &ASAS-SN\\
		2453048.65734	& ASAS	   & 2458055.74717	 &ASAS-SN \\
		2454205.53838	 &ASAS	   &2458244.55000	&ASAS-SN\\
		\hline       
	\end{tabular}
	\vspace{0.25cm}
	\\
\label{Tab: Tab. 1}
\end{table}

\subsection{Multiband WISE and 2MASS photometry}
\label{subsec: 2.3}

\begin{figure}{}
	\begin{center}
		\includegraphics[trim=0.2cm 0.0cm 0.1cm 0.0cm,clip,width=0.45\textwidth,angle=0]{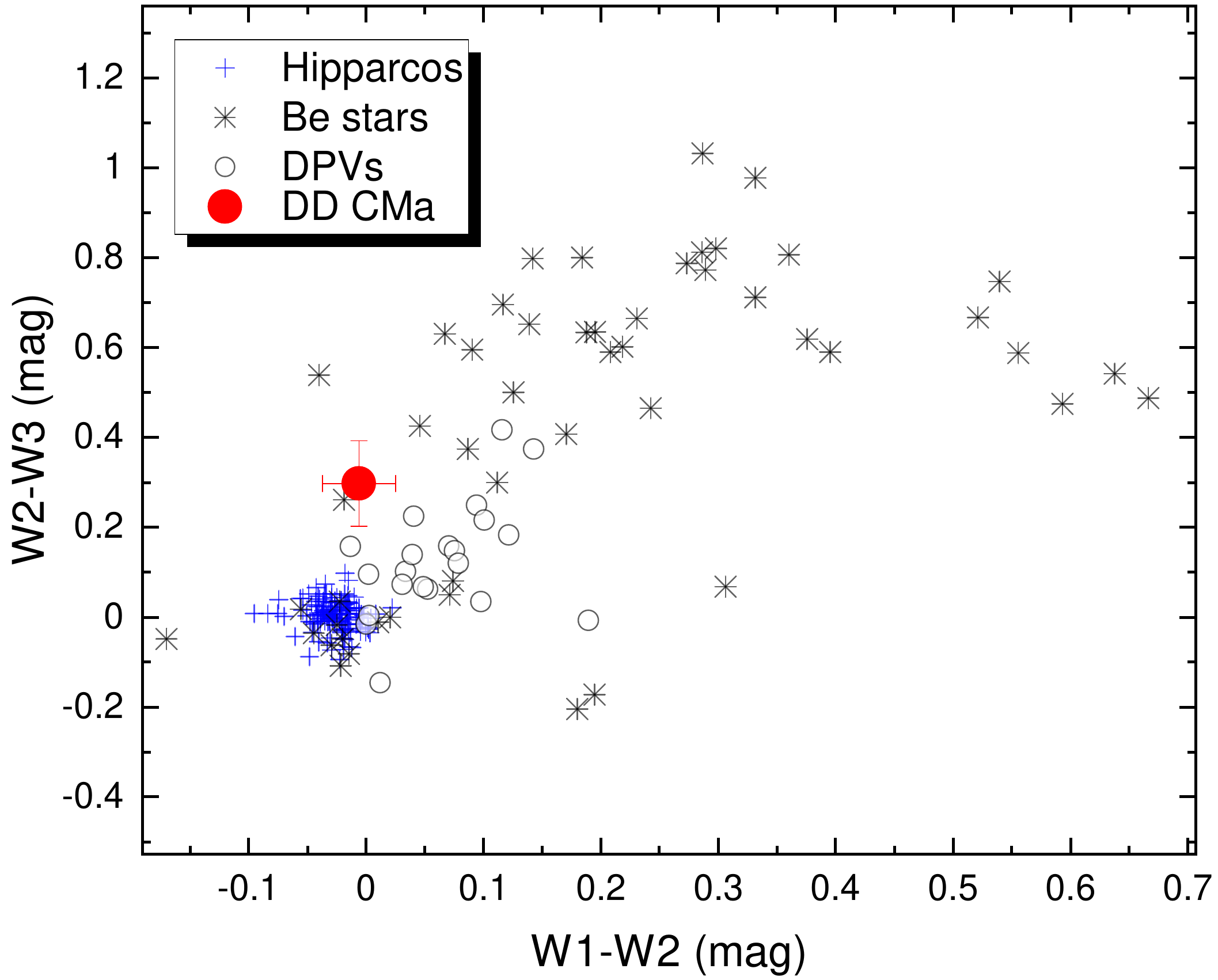}
		\includegraphics[trim=0.2cm 0.0cm 0.0cm 0.1cm,clip,width=0.45\textwidth,angle=0]{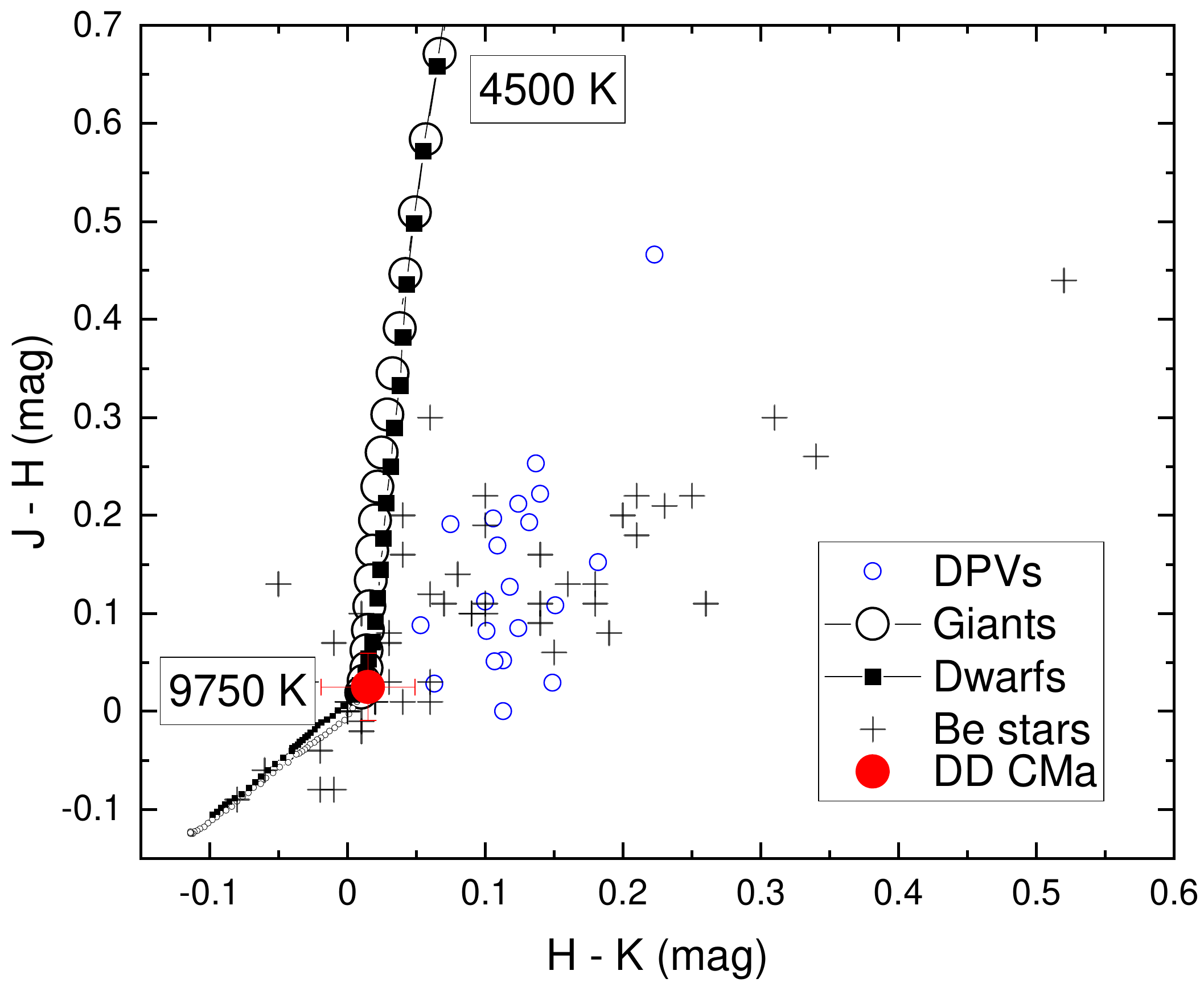}
	\end{center}
	\caption{(Up) $W2-W3$ versus $W1-W2$ color-color diagram for DD\,CMa and systems with and without circumstellar matter. 
		In this diagram the blue crosses represent B1\,V to K3\,V single stars from the Hipparcos catalogue  \citep{1997A&A...323L..49P}. (Down) The 2MASS $J-H$ versus $H-K$ color-color diagram for DD\,CMa 
		and systems with and without circumstellar matter. For details of data selection in both figures see  \citet{2016MNRAS.455.1728M}.}
	\label{fig:Fig. 4}
\end{figure}

\noindent
We performed a search for photometric data in the database of the Wide-field Infrared Survey Explorer (WISE) \citep{2010AJ....140.1868W}  and find mean colors of $W1-W2$= -0.006 $\pm$ 0.031 mag and $W2-W3$= 0.297 $\pm$ 0.095 mag.  In addition, using  data of the Two Micron All Sky Survey (2MASS) \citep{2006AJ....131.1163S} we obtained mean colors  of $J-H$= 0.015 $\pm$ 0.034 mag and $H-K$= -0.025 $\pm$ 0.034 mag. We did not correct the colors for interstellar extinction since it should be insignificant at these  wavelengths. We compared these infrared colors with those of systems with circumstellar envelopes, namely, Be stars - rapidly rotating B-type stars with circumstellar disks -, Double Periodic Variables (DPVs) - close binaries similar to $\beta$ Lyrae showing super-orbital photometric cycles -, and W Serpentis - close interacting binaries showing large variability. The comparison data used in this work are from \citet{2016MNRAS.455.1728M}.
We observe that DD\,CMa does not show color excess in $JHK$ photometry - it is located close to a single star of temperature 10000\,K in the color-color diagram -  but in the WISE color-color diagram the system is located in the area of objects with circumstellar envelopes showing an excess in $W2-W3$. This can be considered as evidence for circumstellar material  (see Fig.\ref{fig:Fig. 4}).

\section{Spectroscopic analysis}
\label{Sec: Sec. 3}

\noindent
In this section we describe our spectroscopic observations and provide results of our study of
radial velocities and spectral decomposition.

\subsection{Spectroscopic observations}
\label{subsec: 3.1}

\noindent
We conducted spectroscopic observations at the Cerro Tololo Interamerican Observatory with the CHIRON spectrograph mounted in the SMARTS 1.5-meter telescope.
We acquired 10 spectra over year 2018 in service mode and with resolving power $R$ = 25\,000 (fiber mode). These spectra have typical signal-to-noise ratio  
$SNR = 73$, measured around the H$\alpha$ line in the continuum (Table \ref{Tab: Tab. 2}). Canonical corrections by flat and bias, extraction to one-dimensional spectrum and wavelength calibration were done with standard \texttt{IRAF} routines \citep{1993ASPC...52..173T}. The spectra were normalized to the continuum and then corrected to velocities in the heliocentric rest of frame. We did not flux calibrated our spectra but this has not effect for the line strength measurements and radial velocities reported in this paper. In addition, we will dub primary and secondary component the more and less massive star; we will show they are the hotter and cooler star; their parameters will be labeled with indexes \textquotedblleft{h}\textquotedblright \,and \textquotedblleft{c}\textquotedblright, respectively. Our observations cover around 60\% of the orbital cycle  in the spectral range 4505$-$6859 \AA{} with relatively good coverage at quadratures.
\ \\
\begin{table}
	\caption{Summary of spectroscopic observations, where N corresponds to the number of spectra with an exposure time of 3600 seconds for a spectral resolution of $R=25000$.
		$\textrm{HJD'= HJD} - 2450000$ is measured at mid-exposure, $\phi_\mathrm{o}$ is the orbital cycle phase calculated according to equation 1.}
	\normalsize
	\begin{center}
		\begin{tabular}{cccccccccc}
			\hline
			\hline
			\noalign{\smallskip}
			\textrm{UT-date}  	&  \textrm{HJD'}		  & \textrm{$\phi_{\mathrm{o}}$} & \textrm{S/N} \\
			\hline
			\textrm{29-01-2018} & \textrm{8148.70016} &  \textrm{0.302} &  \textrm{93}  \\
			\textrm{30-01-2018} & \textrm{8149.64995} &  \textrm{0.775} &  \textrm{74}  \\
			\textrm{12-02-2018} & \textrm{8162.67825} &  \textrm{0.262} &  \textrm{76}  \\
			\textrm{13-02-2018} & \textrm{8163.58789} &  \textrm{0.715} &  \textrm{49}  \\
			\textrm{26-02-2018} & \textrm{8176.57953} &  \textrm{0.184} &  \textrm{87}  \\
			\textrm{27-02-2018} & \textrm{8177.57591} &  \textrm{0.680} &  \textrm{55}  \\
			\textrm{12-03-2018} & \textrm{8190.53966} &  \textrm{0.134} &  \textrm{85}  \\
			\textrm{13-03-2018} & \textrm{8191.52936} &  \textrm{0.627} &  \textrm{59}  \\
			\textrm{26-03-2018} & \textrm{8204.50432} &  \textrm{0.087} &  \textrm{68}  \\
			\textrm{27-03-2018} & \textrm{8205.48964} &  \textrm{0.578} &  \textrm{78}  \\
			\hline
		\end{tabular}
	\end{center}
	\label{Tab: Tab. 2}
\end{table}

\subsection{Radial velocities}
\label{subsec: 3.2}

\noindent
We find sets of lines of both stellar components in the spectra, i.e. the system 
is a SB2 binary. Therefore, we were able to trace the movement of each stellar component individually, by measuring their radial velocities (RVs), 
using different absorption lines. We used for the primary (more massive star) He\,I\,4713, H$\beta$ $\lambda$ 4861.33, He\,I\,4921.9, He\,I\,5875, H$\alpha$ $\lambda$ 6562.817, and He\,I \,6678.149, and for the secondary (less massive star) H$\beta$ $\lambda$ 4861.33 \ and H$\alpha$ $\lambda$ 6562.817.

In order to determine the line profile center, peak strength, equivalent width and radial velocity for these lines, we used the 
deblending routine included in the \texttt{splot} \texttt{IRAF} task. This allowed us to define the continuum region and the
initial positions for each blended line, that were fitted with Gaussian type functions of adjustable position and broadening. 
The measured radial velocities for the secondary and primary component are given in Table \ref{Tab: Tab. 3}. 

\begin{figure}
	\begin{center}
		\includegraphics[trim=0.0cm 0.0cm 0.0cm 0.0cm,clip,width=0.45\textwidth,angle=0]{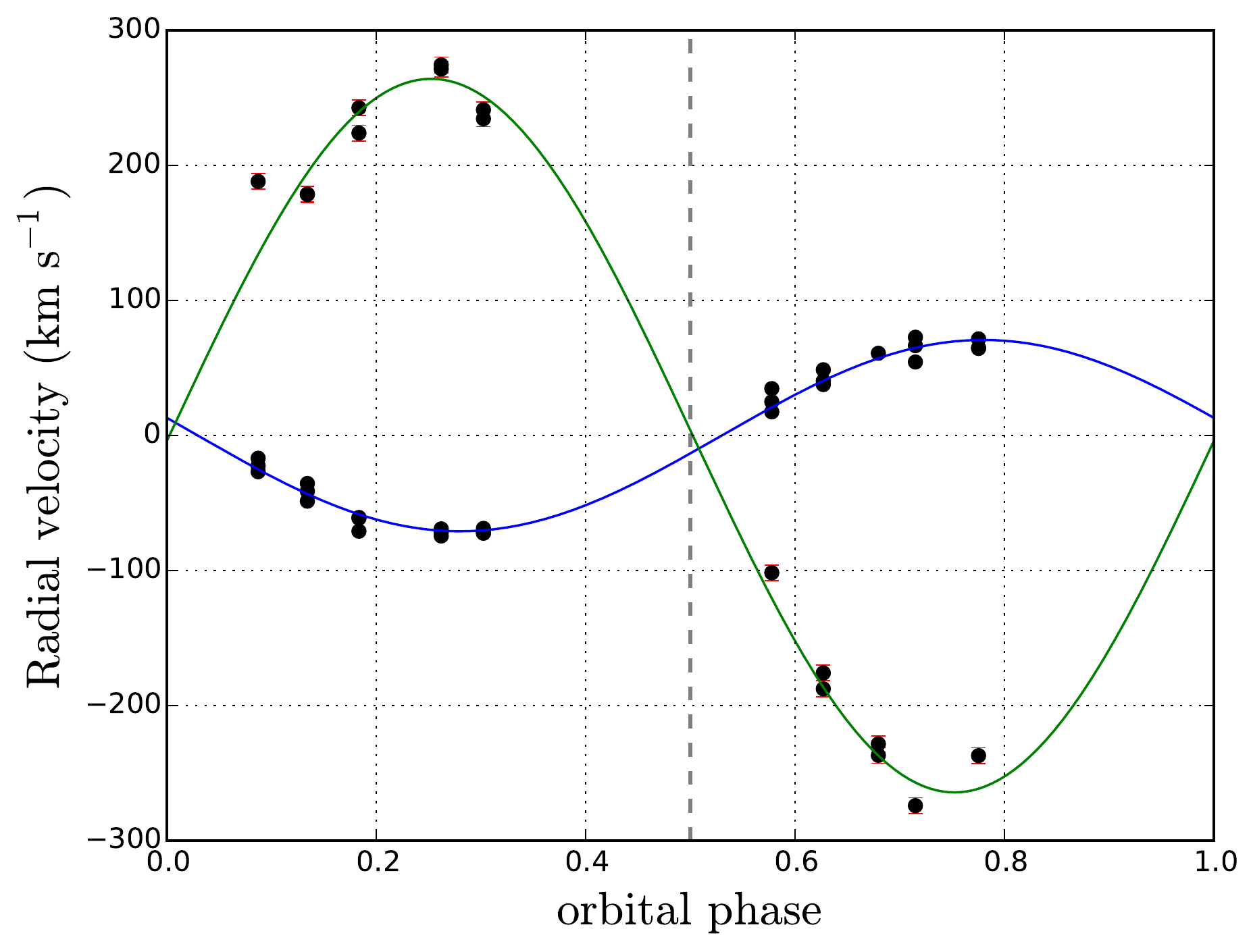}
	\end{center}
	\caption{Radial velocities of DD\,CMa using H$_\beta$ and He\,I lines (4713 and 4921.9 \AA) for the more massive star (points following the blue line) and H$\alpha$ and H$\beta$ lines for the less massive star (points following the green curve).  All measurements were done with Gaussian fits to the spectral lines. Sinusoidal adjustments assuming a circular orbit are shown.}
	\label{fig:Fig. 5}
\end{figure}

The RVs for the secondary were fitted using a sine function through a Marquart-Levenberg method \citep{1963SIAM...11..431} of non-linear least square fit, wherein the solutions and respective errors are obtained through an iterative succession of local linearization. We obtained an amplitude $K_\mathrm{c} = 265.1 \pm 7.3 ~\mathrm{km\,s^{-1}}$ and a zero point $0.5 \pm 5.5 ~\mathrm{km\,s^{-1}}$. In the case of the primary the RVs were fitted using a sine function with $K_\mathrm{h} = 70.9 \pm 1.8 ~\mathrm{km\,s^{-1}}$ and zero point $0.0 \pm 1.1 ~\mathrm{km\,s^{-1}}$. Both solutions assume a circular orbit (Fig. \ref{fig:Fig. 5}).

\begin{table*}
	\caption{Radial velocities and their errors of the less massive star (H$\alpha$  and H$\beta$) and the more massive star (He\,I\,4713 \AA{}, H$\beta$ and He\,I\,4921.9 \AA). We give HJD' = HJD -2\,450\,000. }
	\normalsize
	\begin{center}
		\begin{tabular}{cccccccccccc}
			\hline
			\noalign{\smallskip}
			\textrm{HJD'} 	& \textrm{H$\alpha$ } & \textrm{error } & \textrm{H$\beta$ } 	& \textrm{error}  & \textrm{HeI 4713}& \textrm{error } & \textrm{H$\beta$} & \textrm{error} & \textrm{HeI 4921.9} & \textrm{error}\\               
			\textrm{}	& \textrm{(\kms)}& \textrm{(\kms)}& \textrm{(\kms)}& \textrm{(\kms)} & \textrm{(\kms)}   & \textrm{(\kms)} & \textrm{(\kms)}	& \textrm{(\kms)}& \textrm{(\kms)}& \textrm{(\kms)}\\
			\hline
			\hline
			\textrm{8148.70016} 	& \textrm{ 241.3} 		& \textrm{5.8} 		& \textrm{ 234.6} 		& \textrm{5.8}    & \textrm{-68.7} 	& \textrm{1.2} & \textrm{-71.8}		& \textrm{1.2} & \textrm{-72.3}		& \textrm{1.2}\\
			\textrm{8149.64998} 	& \textrm{-237.0} 		& \textrm{5.8} 		& \textrm{-307.5} 		& \textrm{5.8} & \textrm{ 71.6} 	& \textrm{1.2} & \textrm{ 65.0} 	& \textrm{1.2} & \textrm{ 64.6}		& \textrm{1.2}\\
			\textrm{8162.67824} 	& \textrm{ 271.5} 		& \textrm{5.8} 		& \textrm{ 274.5} 		& \textrm{5.8} & \textrm{-71.8} 	& \textrm{1.2} & \textrm{-69.0} 	& \textrm{1.2} & \textrm{-74.3}		& \textrm{1.2}\\
			\textrm{8163.58789} 	& \textrm{-274.0} 		& \textrm{5.8} 		& \textrm{------} 		& \textrm{---} & \textrm{ 72.8} 	& \textrm{1.2} & \textrm{ 66.6} 	& \textrm{1.2} & \textrm{ 54.5}		& \textrm{1.2}\\
			\textrm{8176.57953} 	& \textrm{ 224.1} 		& \textrm{5.8} 		& \textrm{ 242.7} 		& \textrm{5.8} & \textrm{-60.6} 	& \textrm{1.2} & \textrm{-61.2} 	& \textrm{1.2} & \textrm{-70.8}		& \textrm{1.2}\\
			\textrm{8177.57591} 	& \textrm{-236.8} 		& \textrm{5.8} 		& \textrm{-228.3} 		& \textrm{5.8} & \textrm{-----} 	& \textrm{---} & \textrm{-----} 	& \textrm{1.2} & \textrm{ 61.0}		& \textrm{1.2}\\
			\textrm{8190.53966} 	& \textrm{ 178.8} 		& \textrm{5.8} 		& \textrm{ 178.6} 		& \textrm{5.8} & \textrm{-35.4} 	& \textrm{1.2} & \textrm{-48.5} 	& \textrm{1.2} & \textrm{-41.1}		& \textrm{1.2}\\
			\textrm{8191.52936} 	& \textrm{-175.8} 		& \textrm{5.8} 		& \textrm{-187.5} 		& \textrm{5.8} & \textrm{ 48.8} 	& \textrm{1.2} & \textrm{ 40.4} 	& \textrm{1.2} & \textrm{ 37.8}		& \textrm{1.2}\\
			\textrm{8204.50431} 	& \textrm{ 188.2} 		& \textrm{5.8} 		& \textrm{------} 		& \textrm{---} & \textrm{-22.3} 	& \textrm{1.2} & \textrm{-16.7} 	& \textrm{1.2} & \textrm{-26.8}		& \textrm{1.2}\\
			\textrm{8205.48963} 	& \textrm{-101.6} 		& \textrm{5.8} 		& \textrm{------} 		& \textrm{---} & \textrm{ 34.8} 	& \textrm{1.2} & \textrm{ 17.6} 	& \textrm{1.2} & \textrm{ 25.0}		& \textrm{1.2}\\
			\hline
		\end{tabular}
	\end{center}
\label{Tab: Tab. 3}
\end{table*}

In order to obtain the orbital parameters of DD\,CMa we used a public subroutine based on a genetic algorithm called {\sc pikaia} \citep{1995ApJS..101..309C}. The code determines the single parameter set that minimizes the difference between the model\textquoteright{s} predictions and the data, producing a series of theoretical velocities and finding the best parameters through a minimization of chi-square:

\small
\begin{equation}
\chi^2_6(P_\mathrm{o},\tau,\omega,e,K,\gamma)
= \frac{1}{N-6}\sum_\mathrm{j=1}^{N} \left(\frac{V_\mathrm{j}^\mathrm{obs}-V(t_\mathrm{j};P_\mathrm{o},\tau,\omega,e,K,\gamma)}{\sigma_\mathrm{j}}  \right)^2,
\label{eq: eq. 2}
\end{equation}

\normalsize
\noindent
wherein $(N - 6)$ corresponds to the number of degrees of freedom of the fit to 6 parameters, $N$ is the number of observations, $V^\mathrm{obs}_\mathrm{j}$ is the observed radial velocity in the dataset, and $V(t_\mathrm{j}; P_\mathrm{o}, \omega, e, K, \gamma)$ is the radial velocity at the time $t_\mathrm{j}$ . The other parameters such as $P_\mathrm{o}$ represent the orbital period, $\tau$ the time of passaging through the periastron, $\omega$  the periastron longitude, $e$ the orbital eccentricity, $K$ the half-amplitude of the radial velocity and $\gamma$ the velocity of the system center of mass. The theoretical radial velocities are computed through the equation 2.45 given by \citet{2001icbs.book.....H}:
\begin{equation}
V(t)= \gamma + K (\cos(\omega +\theta(t))+e\cos(\omega)),
\label{eq: eq. 3}
\end{equation}

\noindent
in order to solve the equation \ref{eq: eq. 3} it is necessary to compute  the true anomaly $\theta$. 
This parameter is obtained after solving the equation relating the true and the eccentric anomaly: 
\begin{equation}
\tan\left(\frac{\theta}{2}\right)= \sqrt{\frac{1+e}{1-e}}\tan\left(\frac{E}{2}\right),
\label{eq: eq. 4}
\end{equation}

\noindent
but to obtain the value of eccentric anomaly $E$ it is necessary to solve the equation 2.35 of \citet{2001icbs.book.....H}: 
\begin{equation}
E-e\sin(E)=\frac{2\pi}{P_{\mathrm{o}}}(t-\tau),
\label{eq: eq. 5}
\end{equation}

\noindent
since the equation \ref{eq: eq. 5} has not analytic solution, it is necessary to use an iterative method to solve it. 
Then, after computing the solution, we can solve the equation \ref{eq: eq. 3} obtaining  the radial velocity solution. 
Following this procedure we obtained the orbital parameters showed in Table \ref{Tab: Tab. 4}.

In addition, we implemented a test for the significance of the eccentricity using the following equation \citep{1971AJ.....76..544L}: 
\begin{equation}
p_{1}= \left(\frac{\sum(O-C)^{2}_{ecc}}{\sum(O-C)^{2}_{circ}}\right)^{(n-m)/2}
\label{eq: eq. 6}
\end{equation}

\begin{figure*}
	\begin{center}
		\includegraphics[trim=0.0cm 0.0cm 0.0cm 0.0cm,clip,width=0.3\textwidth,angle=0]{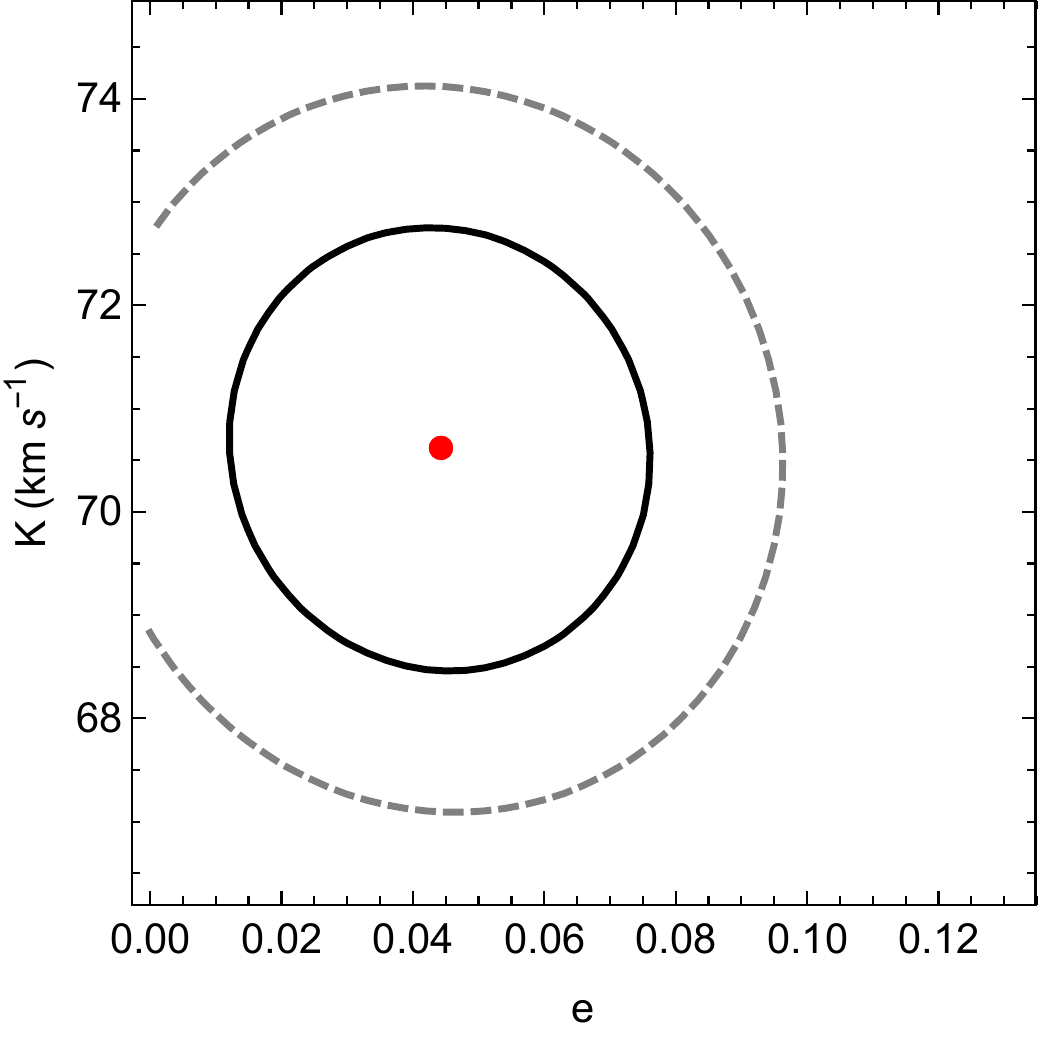}
		\includegraphics[trim=0.0cm 0.0cm 0.0cm 0.0cm,clip,width=0.3\textwidth,angle=0]{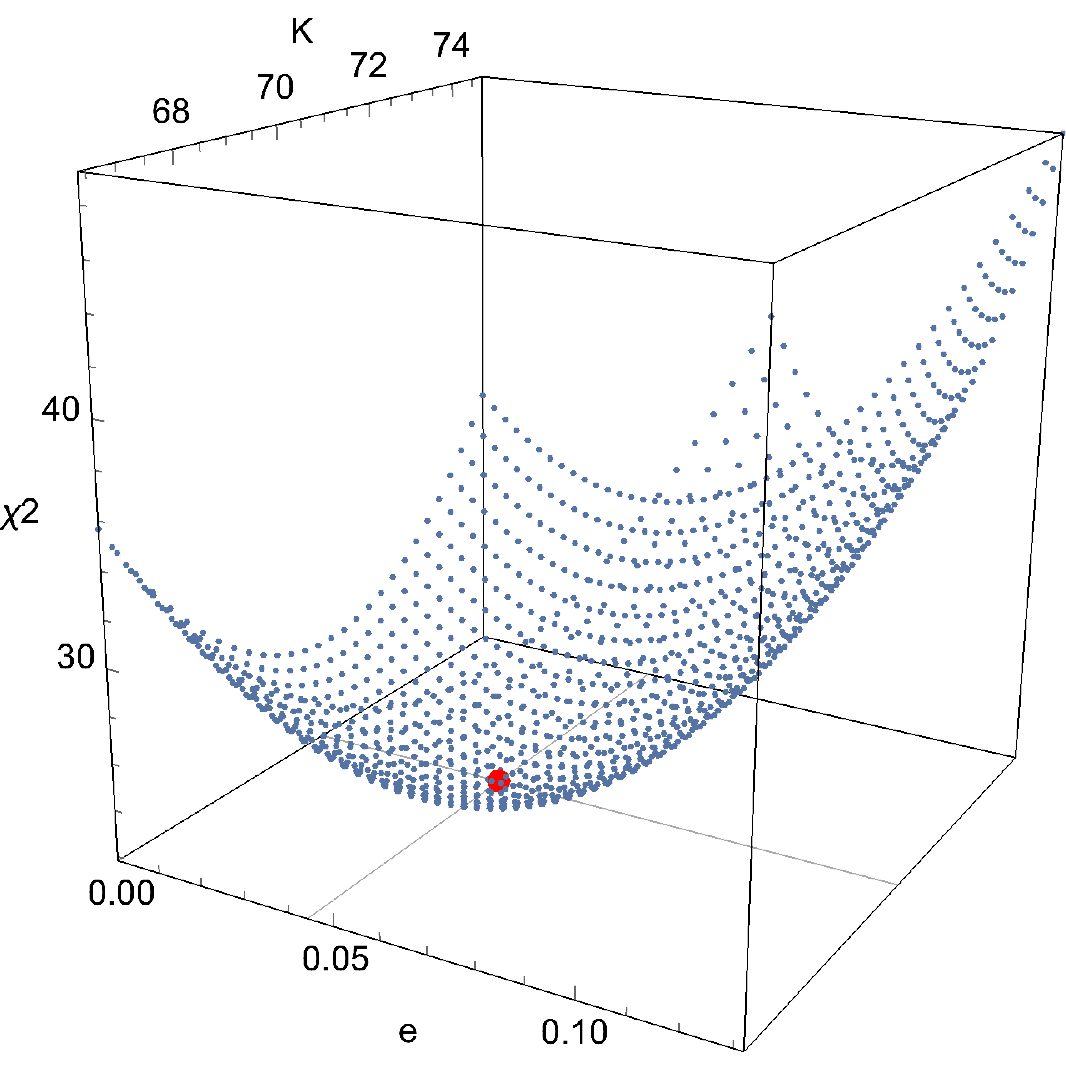}	
		\includegraphics[trim=0.0cm 0.0cm 0.0cm 0.0cm,clip,width=0.39\textwidth,angle=0]{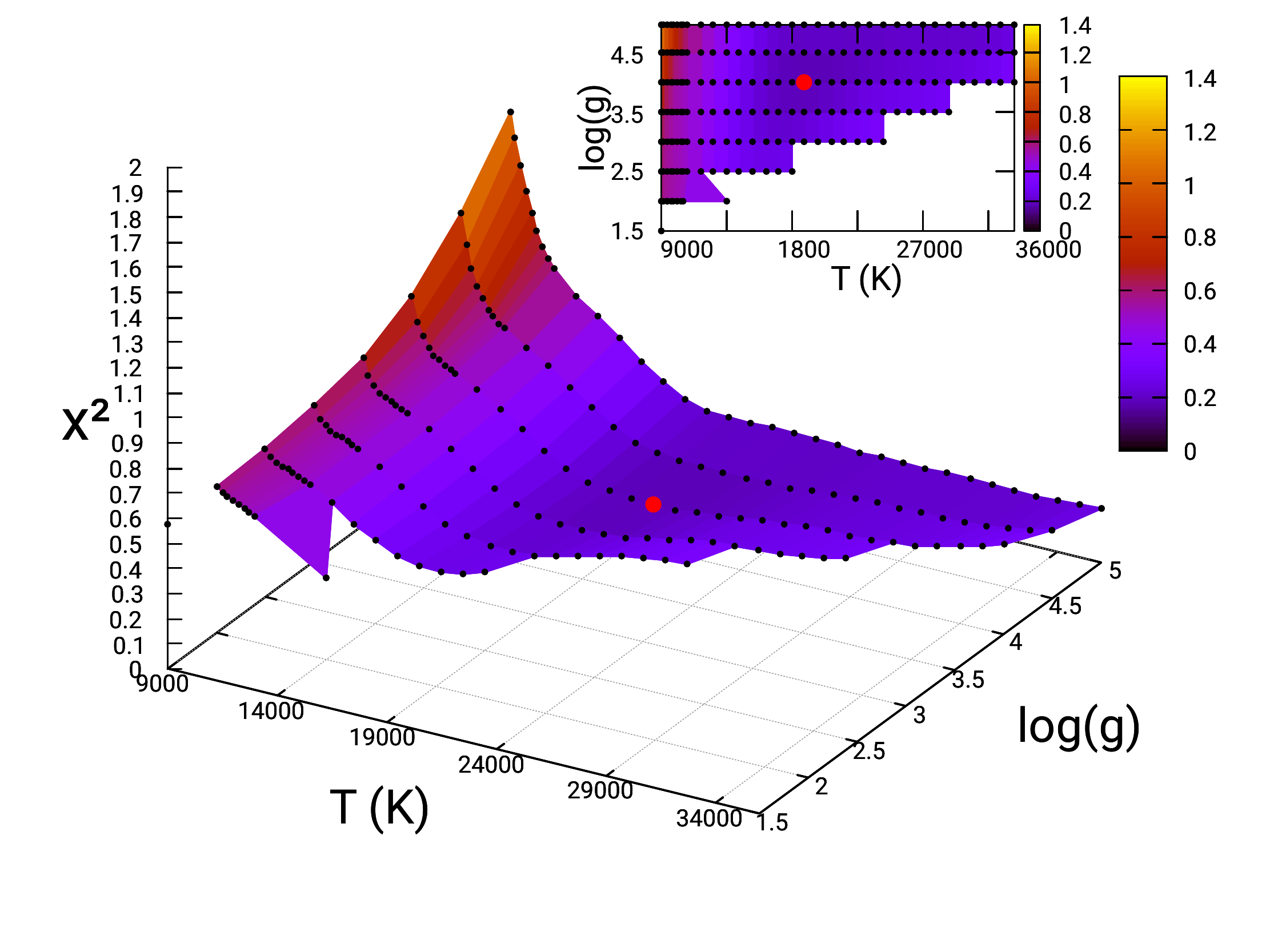}
	\end{center}
	\caption{Left: Representation of $\chi^2$ distribution in the K-e plane, calculated through  Monte Carlo Simulations. The solid-black and dashed-gray lines correspond to $1\sigma$ and $2\sigma$ iso-contours, while the red point indicates the minimum $\chi^2$ solution. Middle: Contour corresponding to the $\Delta\chi^2$ that includes $68.3\%$ of the probability. Right: Surface color map of $\chi^2$ analysis obtained for the hotter star. The best model is obtained at $\chi^2= 0.174$ and is represented by a red dot at $T= 20000 ~\mathrm{K}$ and $\log{g} = 4.0 ~\mathrm{dex}$.}
	\label{fig:Fig. 6}
\end{figure*}

\noindent
Lucy's test condition says  that if $p_{1} < 0.05$ an elliptic orbit is accepted, while if $p_{1} \geq 0.05$ a circular orbit is statistically preferred.
The subindex $ecc$ indicates the residuals of an eccentric fit, $circ$ corresponds to the residuals of a circular orbit fit, $n$ is the number of observations and $m$ the number of free parameters used in the eccentric fit.  We obtained $p_{1}= 0.2226$, i.e. DD\,CMa has an orbit compatible with a circular orbit (Fig. \ref{fig:Fig. 6}). We notice consistent results for the orbital solutions obtained from the radial velocities of  the primary and secondary stars. In the next sections we use the orbital period derived from the photometric analysis and the orbital parameters of Table \ref{Tab: Tab. 4}.

\ 

\ \\
\begin{table}
\caption{Orbital parameters for the stellar components of DD\,CMa obtained through minimization of $\chi^2$ given by equation \ref{eq: eq. 4}. The value $\tau^{*}=\tau-2450000$ is given and the maximum and minimum are one isophote of 1$\sigma$. $S_h$ and $S_c$ stand for solutions for the more massive and less massive star, respectively.}
\normalsize
\resizebox{0.5\textwidth}{3.3cm}{
\begin{tabular}{c|rrrr}
\hline
\noalign{\smallskip}
\textrm{}		&	\textrm{Parameter} 				& \textrm{Best Value} & \textrm{Lower limit} & \textrm{Upper limit} \\
\hline
\hline
\textrm{} 		&	\textrm{$\tau^{*}$} 			& \textrm{150.709} 	& \textrm{150.694} 	& \textrm{150.722} 	\\
\textrm{} 		&	\textrm{e} 						& \textrm{0.04419} 	& \textrm{0.02000} 	& \textrm{0.07000} 	\\
\textrm{} 		&	\textrm{$\omega \textrm{(rad)}$}& \textrm{3.31212} 	& \textrm{3.27151} 	& \textrm{3.25651} 	\\
\textrm{$S_\textrm{h}$} &	\textrm{$K_\textrm{h}$ (\kms)} 		& \textrm{70.6} 	& \textrm{68.4} 	& \textrm{72.9} 	\\
\textrm{} 		&	\textrm{$\gamma$ (\kms)} 		& \textrm{0.4} 		& \textrm{-1.2}		& \textrm{2.0} 		\\
\textrm{} 		&	\textrm{N} 						& \textrm{28} 		& \textrm{} 		& \textrm{} 		\\
\textrm{} 		&	\textrm{$\chi^2$} 				& \textrm{25.1513} 	& \textrm{} 		& \textrm{} 		\\
\noalign{\smallskip}
\hline
\noalign{\smallskip}
\textrm{} 		&	\textrm{$\tau^{*}$} 			& \textrm{150.709} 	& \textrm{150.686}  & \textrm{150.730}	\\
\textrm{} 		&	\textrm{e} 						& \textrm{0.00000} 	& \textrm{0.00000}  & \textrm{0.02000}	\\
\textrm{} 		&	\textrm{$\omega \textrm{(rad)}$}& \textrm{0.31410} 	& \textrm{0.24846}  & \textrm{0.38346}	\\
\textrm{$S_\textrm{c}$}&	\textrm{$K_\textrm{c}$ (\kms)} 		& \textrm{264.2} 	& \textrm{253.7} 	& \textrm{275.2}	\\
\textrm{} 		&	\textrm{$\gamma$ (\kms)} 		& \textrm{0.0} 		& \textrm{-8.3}		& \textrm{8.2}	\\
\textrm{} 		&	\textrm{N} 						& \textrm{17} 		& \textrm{} 		& \textrm{}			\\
\textrm{} 		&	\textrm{$\chi^2$} 				& \textrm{14.3765} 	& \textrm{} 		& \textrm{}			\\
\noalign{\smallskip}
\hline
\end{tabular}
}
\label{Tab: Tab. 4}
\end{table}

\subsection{Spectral  decomposition}
\label{subsec: 3.3}

\noindent
We performed the spectral  decomposition using the iterative method proposed by \citet{2006A&A...448..283G}, which allows to calculate the spectra and RVs of two stellar components of a binary system, using alternately the spectrum of one component to calculate the spectrum of the other. The process eliminates gradually in every step the spectral features of one stellar component.
The method is executed iteratively until convergence is assured, i.e. until the flux contribution of one component virtually disappears.
As input parameters we used the theoretical radial velocities obtained from the sinusoidal fits in the previous section and performed seven iterations for both components, obtaining successfully clean average spectra for both stars.

\subsection{Physical parameters of the more massive star and interstellar reddening}
\label{subsec: 3.4}

We have compared the  decomposed average spectrum of the hotter star with a grid of synthetic spectra constructed with the stellar spectral synthesis program  {\textsc{spectrum}} \footnote{\url{http://www.appstate.edu/~grayro/spectrum/spectrum.html}}, which uses atmospheric models computed of ATLAS9\footnote{\url{http://wwwuser.oats.inaf.it/castelli/grids.html}} \citep{2003IAUS..210P.A20C} in Local Thermodynamic Equilibrium (LTE). We looked for the best fit synthetic spectrum by minimizing residuals among the observed and the theoretical spectra, considering the veiling factor. Briefly, this factor is a correction incorporated on the spectrum of the hotter star that takes into account the additional light that is contributed by the companion. This veiling produces an artificial weakness of the absorption lines that can be removed using the aforementioned factor. 

The theoretical models were computed using 2 groups of effective temperatures, the first considers $9000 \leq T_\mathrm{h} \leq 11000 ~\mathrm{K}$ with steps of 250 K and the second group considers $11000 \leq T_\mathrm{h} \leq 36000 ~\mathrm{K}$ with steps of 1000 K. This difference between the resolution of both groups matches the grid resolution.  The surface gravity varies from 1.5 to 5.0 dex with steps of 0.5 dex, the macro-turbulent velocity varies from 0 to 10 ~$\mathrm{km\,s^{-1}}$ with steps of 1 ~$\mathrm{km\,s^{-1}}$, the $v\sin{i}$ varies from 10 to 140 ~$\mathrm{km\,s^{-1}}$ with steps of 10 ~$\mathrm{km\,s^{-1}}$ and we considered two values for the micro-turbulent velocity of 0.0 and 2.0 ~$\mathrm{km\,s^{-1}}$.
The following parameters are kept fixed in the grid: the stellar metallicity $M/H$=0  and the mixing length parameter $l/H$=1.25. In addition, we
constructed spectra with veiling factor $\eta$ running 0.0 to 0.9 with step 0.1.

\begin{figure}
	\begin{center}
		\includegraphics[trim=0.0cm 0.0cm 0.0cm 0.0cm,clip,width=0.45\textwidth,angle=0]{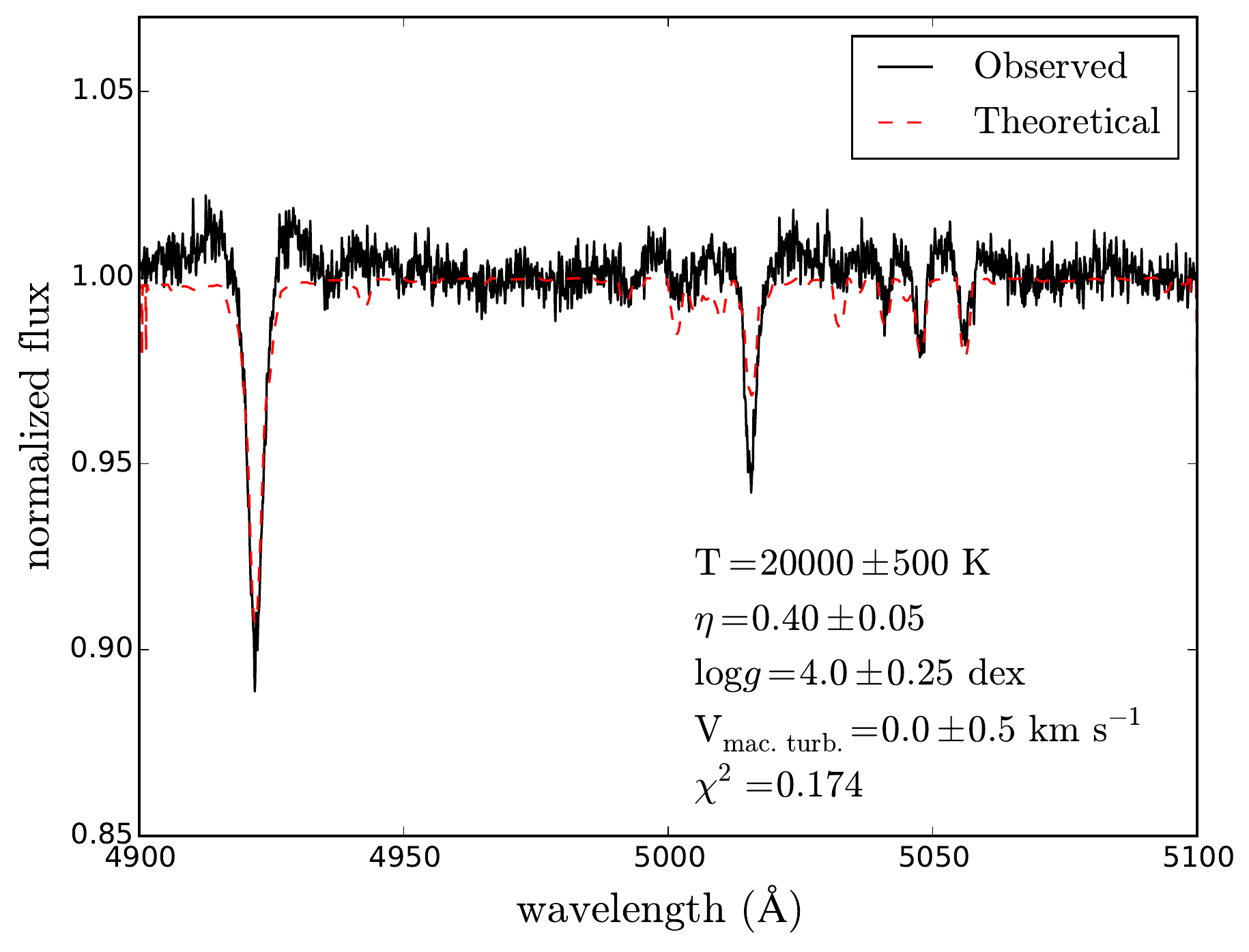}
	\end{center}
	\caption{Detailed comparison between the  decomposed observed (black line) and theoretical (red line) primary spectrum.}
	\label{fig:Fig. 7}
\end{figure}

The implemented method converged successfully to a minimum chi-square at $T_\mathrm{h} = 20000 \pm 500 ~\mathrm{K}$,  $\log{g_\mathrm{h}} = 4.0 \pm 0.25 ~\mathrm{dex}$, $v_\mathrm{mac} = 1.0 \pm 0.5 ~\mathrm{km\,s^{-1}}$, $v_\mathrm{mic} = 0.0$, $v_\mathrm{1r} \sin{i} = 80 \pm 5 ~\mathrm{km\,s^{-1}}$ and $\eta = 0.4 \pm 0.05$ with six freedom degrees and a best value of $\chi^2_6 = 0.174$ (Fig. \ref{fig:Fig. 6}{-Right} and Fig. \ref{fig:Fig. 7}).

In addition, we measured the equivalent width (EW) of the diffuse interstellar bands (DIBs) of every spectrum by fitting simple Gaussians to the 5780 \AA{} and 5797 \AA{} lines to determine the average reddening in the system's direction, that will be used later to determine the distance of the system. For that, we have used the formula given by \citet{1993ApJ...407..142H} computing an average reddening of $E(B-V)= 0.46 \pm 0.03$ and an interstellar absorption $A_\mathrm{V}= 1.42 \pm 0.09$.

\subsection{Mass ratio, Roche lobe radius and structure}
\label{subsec: 3.5}

\noindent
After exploring the spectrum of the hotter star we noticed that it reveals emission in H$\alpha$ and H$\beta$ that are characteristics of interacting binaries of the 
Algol-type. Therefore, we assume a semidetached nature of the system, i.e.
the less massive star fills its Roche lobe. From the radial velocity half-amplitudes we get a system mass ratio $q = K_\mathrm{h}/K_\mathrm{c}= 0.268 \pm 0.01$.  On the other hand, we obtain the ratio between the effective Roche lobe radius of the less massive star  $R_\mathrm{L}$ and the orbital separation $a$, using the following relation by \citet{1983ApJ...268..368E}:

\begin{equation}
\frac{R_\mathrm{L}}{a}= \frac{0.49q^{2/3}}{0.6q^{2/3}+ \ln(1+q^{1/3})},
\label{eq: eq. 7}
\end{equation}

\noindent
Assuming $R_\mathrm{c}= R_\mathrm{L}$, the above equation gives us $R_\mathrm{c}/a = 0.274 \pm 0.008$. Now, if a star fills its Roche Lobe, its mean density $\bar{\rho}$ is directly related to its orbital period \citep[e.g.][]{1983ApJ...268..368E,2006epbm.book.....E}, and using  $\bar{\rho_c}= 3M_\mathrm{c}/4\pi R^{3}_{\mathrm{c}}$, $M_{\mathrm{c}}=Mq/(1+q)$ along with the Kepler's third law, it is possible to obtain:
\begin{equation}
\bar{\rho_c}= \frac{3q}{(1+q)}\frac{1}{(R_{\mathrm{ 2}}/a)^{3}}\frac{\pi}{GP^{2}}=  0.048 \pm 0.004 ~\mathrm{g\,cm^{-3}}
\label{eq: eq. 8}
\end{equation} 

\noindent 
thus the obtained value is close to that expected for a giant of B9 type. The derivation of the spectral types of the binary stellar components is given in Section \ref{Sec: Sec. 7}.


\section{Light curve model and system parameters}
\label{Sec: Sec. 4}

\noindent
In order to extract the physical information hidden in the orbital light curve we use an algorithm that fit 
the data solving the inverse problem  \citep{1992Ap&SS.197...17D,1996Ap&SS.243..413D,2013AJ....145...80D}. Briefly, this program is based on the Roche model and the principles described by \citet{1971ApJ...166..605W}. The program determines the optimal stellar and physical parameters that best fit the observed and theoretical light curve, through an iterative cycle of corrections, solving the inverse problem with a method based on the one described by \citet{1963SIAM...11..431}. 
We assume synchronous rotation for both components. This is justified since the stars are quite close, so tidal forces are expected to have synchronized the stellar spins rapidly. 
In addition, $q$ and $T_\mathrm{h}$ are fixed to the values determined in our previous spectroscopic study.

The results of the model of the ASAS $V$-band light curve are give in Table \ref{Tab: Tab. 5}-left.
The binary consists of a hot star of mass  $M_\mathrm{h} = 6.3 \pm 0.1 ~\mathrm{M_{\odot}}$, temperature $T_\mathrm{h} = 20000 \pm 500 ~\mathrm{K}$ and radius $R_\mathrm{h} = 3.7 \pm 0.1 ~\mathrm{R_\mathrm{\odot}}$
while the most evolved star has a mass $M_\mathrm{c} = 1.7 \pm 0.1 ~\mathrm{M_{\odot}}$, a temperature $T_{\mathrm{c}} = 11320 \pm 200 ~\mathrm{K}$ and radius $R_\mathrm{c} = 3.6 \pm 0.1 ~\mathrm{R_{\odot}}$. Both stars are separated by $13.4 \pm 0.2 ~\mathrm{R_{\odot}}$. The orbital inclination is $84.\!\!^{\circ}0 \pm 0.\!\!^{\circ}3$.

\begin{figure*}
	\includegraphics[]{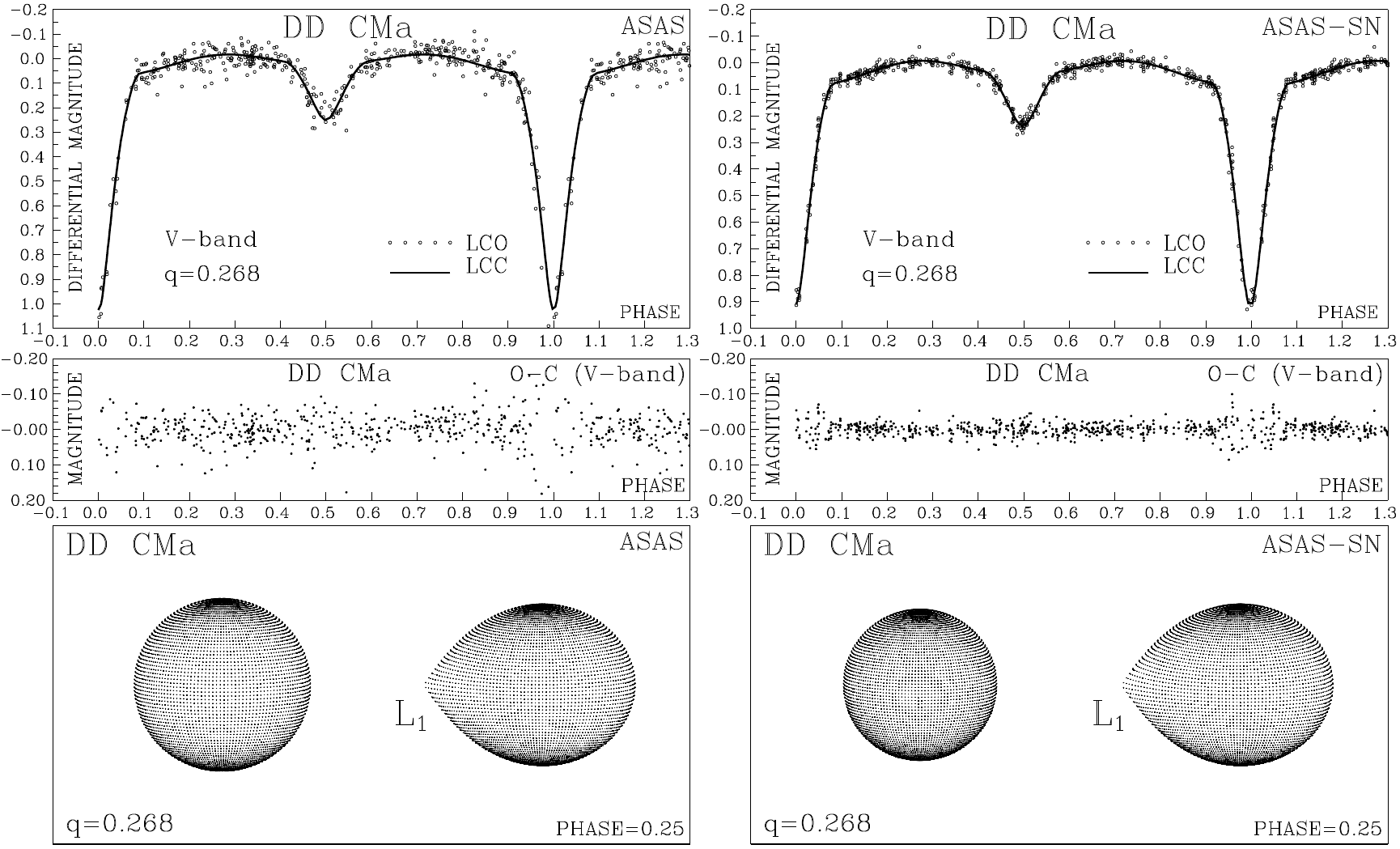}
	\caption{Observed (LCO), synthetic (LCC) light-curves and  the final O-C residuals between the observed and synthetic light curves, and views of the model at orbital phase 0.25, obtained with parameters derived by light curve analysis.}
	\label{fDDCMa}
	\label{fig:Fig. 8}	
\end{figure*}

The results of the model of the ASAS-SN $V$-band light curve are give in Table \ref{Tab: Tab. 5}-right and they are in general in close agreement with those obtained with ASAS data.
We notice the following differences: ASAS-SN $R_\mathrm{h} = 3.2 \pm 0.1 ~\mathrm{R_\mathrm{\odot}}$, i.e. it is $0.5 \pm 0.1 ~\mathrm{R_{\odot}}$ smaller than the value obtained with ASAS, and the ASAS-SN orbital inclination of $81.\!\!^{\circ}0 \pm 0.\!\!^{\circ}2$, which has a variation of $2.\!\!^{\circ}2 \pm 0.\!\!^{\circ}4 $ regarding the ASAS value. These variations are larger than the formal errors, however the observed changes are probably influenced by the different quality of the data (ASAS light curve is noisier) and it is possible that some of errors are underestimated.
Geometrical views along with the observed and calculated light curves for the ASAS and ASAS-SN model are given in Fig. \ref{fig:Fig. 8}. The interesting fact that the scatter of the residuals is larger on the primary eclipse might indicate that our simple model of two stars is not able to represent a more complex system, with eventually additional emitting /absorbing light sources, as expected in a mass-transferring system, as discussed in Sec \ref{Sec: Sec. 7}. At present, we have no  information about structure of the circumstellar matter neither gas stream, hence it is not possible to include these additional emission/absorption structures  in the light curve model.  Therefore,  we caution about a possible bias in the orbital and stellar parameters derived from the light curve model, because of the modeling limitations.

\ \\
\begin{table}
	\caption{(Left) Results of the analysis of DD\,CMa ASAS (Left) and ASAS-SN (Right) V-band light-curve obtained by solving the inverse problem for the Roche model.}
	\label{DD CMa}
	\normalsize
	\resizebox{0.45\textwidth}{6cm}{$
		\begin{array}{llll}
		\textrm{ASAS} 						& \textrm{} 		&  \textrm{ASAS-SN} 						& \textrm{} \\
		\hline
		\noalign{\smallskip}
		\textrm{Quantity} 					& \textrm{} 		& \textrm{Quantity} 						& \textrm{} \\
		\hline
		\noalign{\smallskip}
		n 									& 448				&  n 										& 562	  \\
		{\rm \Sigma(O-C)^2}					& 0.7996 			&  {\rm \Sigma(O-C)^2}						& 0.2230  \\
		{\rm \sigma_{rms}}					& 0.0423			&  {\rm \sigma_{rms}}						& 0.0199  \\
		i {\rm [^{\circ}]}          		& 84.0 \pm 0.3		&  i {\rm [^{\circ}]}                      	& 81.8 \pm 0.2 \\
		{\rm F_h}                   		& 0.584\pm  0.01	&{\rm F_h}                               	& 0.510\pm 0.009\\ 
		{\rm F_c}							& 1.000\pm  0.003	& {\rm F_c}								& 1.000\pm 0.002\\
		{\rm T_h} [{\rm K}]					& 20000 			& {\rm T_h} [{\rm K}]						& 20000 \\
		{\rm T_c} [{\rm K}]					& 11320 \pm 200 	& {\rm T_c} [{\rm K}]						& 11350 \pm 100 \\
		{\rm \Omega_h}						& 3.944  			& {\rm \Omega_h}							& 4.478  \\
		{\rm \Omega_c}						& 2.394  			&  {\rm \Omega_c}							& 2.394  \\
		R_{\rm h} [D=1]                     &  0.271			&   R_{\rm h} [D=1]                         &  0.237 \\
		R_{\rm c} [D=1]                     &  0.253    		&  R_{\rm c} [D=1]                         &  0.253 \\
		\cal M_{\rm_h} {[\cal M_{\odot}]}	& 6.3 \pm 0.1 		&   \cal M_{\rm_h} {[\cal M_{\odot}]}		& 6.4  \pm 0.1\\
		\cal M_{\rm_c} {[\cal M_{\odot}]}	& 1.7 \pm 0.1		&   \cal M_{\rm_c} {[\cal M_{\odot}]}		& 1.7  \pm 0.1\\
		\cal R_{\rm_h} {\rm [R_{\odot}]} 	& 3.7 \pm 0.1		&  \cal R_{\rm_h} {\rm [R_{\odot}]} 		& 3.2  \pm 0.1 \\
		\cal R_{\rm_c} {\rm [R_{\odot}]}  	& 3.6 \pm 0.1 		&  \cal R_{\rm_c} {\rm [R_{\odot}]}  		& 3.7  \pm 0.1 \\
		{\rm log} \ g_{\rm_h}		 		& 4.11 \pm 0.02		&  {\rm log} \ g_{\rm_h}		 			& 4.23 \pm 0.02\\
		{\rm log} \ g_{\rm_c}	         	& 3.55 \pm 0.02		&  {\rm log} \ g_{\rm_c}	         		& 3.55 \pm 0.02 \\
		M^{\rm h}_{\rm bol}	         		&-3.44 \pm 0.1		&  M^{\rm h}_{\rm bol}	         			&-3.14 \pm 0.1\\
		M^{\rm c}_{\rm bol}		 			& -0.94\pm 0.1		& M^{\rm c}_{\rm bol}		 				&-0.97 \pm 0.1\\
		a_{\rm orb}  {\rm [R_{\odot}]}      & 13.4 \pm 0.2		& a_{\rm orb}  {\rm [R_{\odot}]}          & 13.5 \pm 0.2 \\ 
		\noalign{\smallskip} 
		\hline
		\noalign{\smallskip} 
		\hline 
		\noalign{\smallskip}
		\end{array}
		$}
	
	\bigskip
	FIXED PARAMETERS: $q={\cal M}_{\rm c}/{\cal M}_{\rm h}=0.268$ - mass ratio of the components, ${\rm T_h=20000 K}$  - temperature of the more-massive hotter
	component (h), ${\rm F_c}=1.0$ - filling factor for the critical Roche lobe of the less massive cooler component (c), $f{\rm _h}=1.0 ; f{\rm _c}=1.00$ - non-synchronous rotation coefficients of the hotter and colder star respectively, ${\rm \beta_h=0.25 ; \beta_c=0.25}$ - gravity-darkening coefficients of the hotter and colder star, ${\rm A_h=1.0 ;A_c=1.0 }$  - albedo coefficients of the hotter and colder star.
	
	\smallskip 
	\noindent 
	Quantities: $n$ - number of observations, ${\rm \Sigma (O-C)^2}$ - final sum of squares of residuals between observed (LCO) and synthetic (LCC) light-curves, ${\rm \sigma_{rms}}$ - root-mean-square of the residuals, $i$ - orbit inclination (in arc degrees), ${\rm F_h}=R_h/R_{zc}$ - filling factor for the critical Roche lobe of the hotter, more-massive star (ratio of the stellar polar radius to the critical Roche lobe radius along z-axis), ${\rm T_c}$ - temperature of the less-massive (cooler), ${\rm \Omega_{h,c}}$ - dimensionless surface potentials of the hotter and cooler star, $R_{\rm h,c}$ - polar radii of the components in units of the distance between their centers; ${\rm L_h/(L_h+L_c)}$ - luminosity {\rm (V-band)} of the more massive, hotter component, $\cal M_{\rm_{h,c}} {[\cal M_{\odot}]}$, $\cal R_{\rm_{h,c}} {\rm [R_{\odot}]}$ - stellar masses and mean radii of stars in solar units, ${\rm log} \ g_{\rm_{h,c}}$ - logarithm (base 10) of the system components effective gravity, $M^{\rm {h,c}}_{\rm bol}$ - absolute stellar bolometric magnitudes, $a_{\rm orb}$ ${\rm [R_{\odot}]}$ - orbital semimajor axis given in the solar radius units.
	\label{Tab: Tab. 5}
\end{table}

\ \\

\section{Spectral Energy Distribution (SED)}
\label{Sec: Sec. 5}

\begin{figure}
	\includegraphics[trim=0.0cm 0.30cm 0.4cm 0.0cm,clip,width=0.45\textwidth,angle=0]{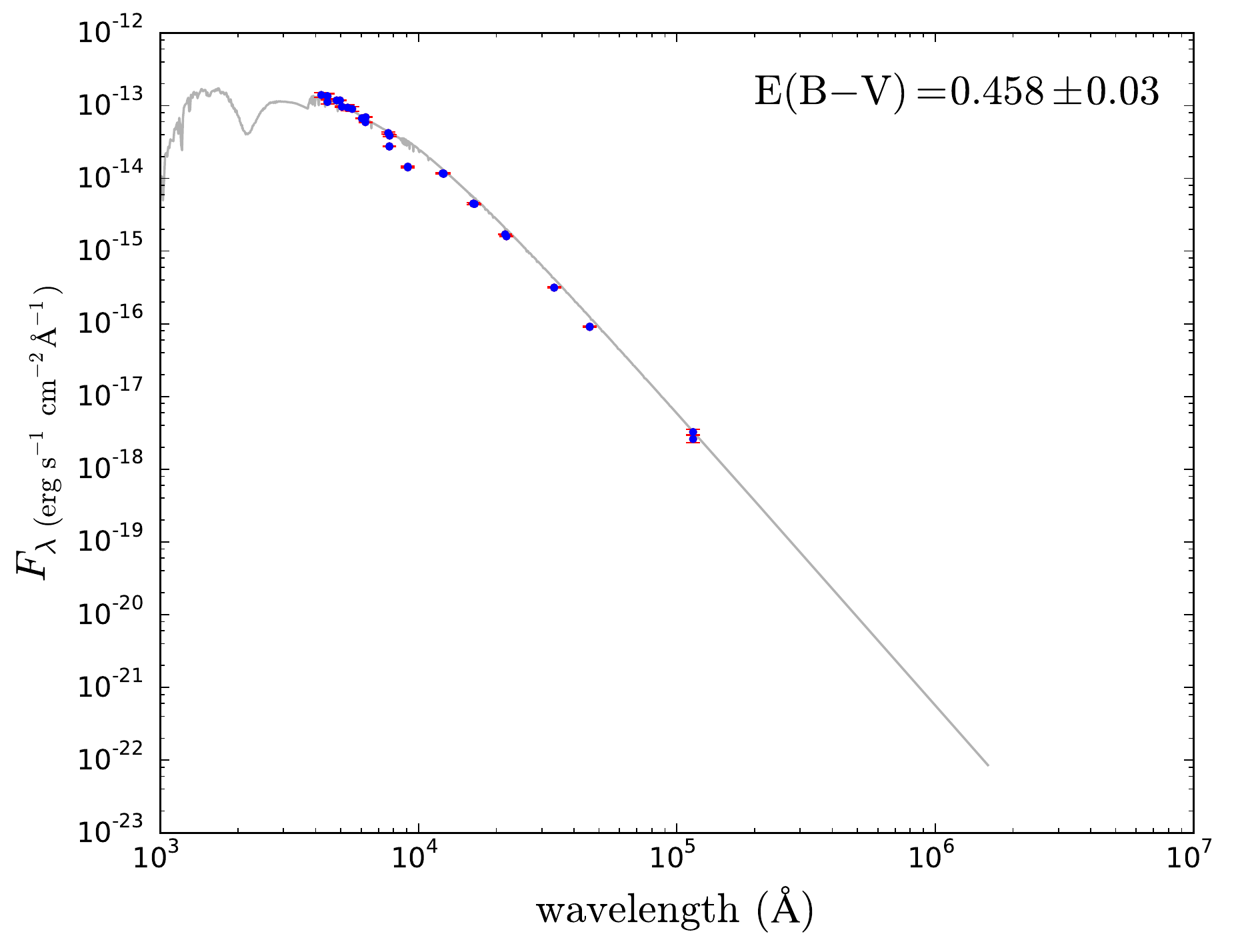}
	\caption{Observed fluxes as listed in Table\,5 (blue dots) along with the best theoretical fit given by Eq.\,10 (solid line).} 
	\label{fig:Fig. 9}
\end{figure}

\noindent
We compiled the photometric fluxes available for DD\,CMa to build the Spectral Energy Distribution, extracting the information from the VizierR Photometric viewer \footnote{\url{http://vizier.u-strasbg.fr/vizier/sed/}}  (Table \ref{Tab: Tab. 6}). 
We selected single star theoretical models from the Virtual Observatory SED Analyzer \footnote{\url{http://svo2.cab.inta-csic.es/theory/vosa/}} \citep{2008A&A...492..277B} to perform a SED fit determining the system distance. 
Specifically, we use those models with $T_{\mathrm{c}}= 11000 ~\mathrm{K}$, $\log{g_{\mathrm{c}}}= 4.0 ~{\mathrm{dex}}$, $T_{\mathrm{h}}= 19000 ~\mathrm{K}$, $\log{g_{\mathrm{h}}}= 3.5 ~{\mathrm{dex}}$. In addition, we considered 
$E(B-V)$ = 0.458 as derived in Section \ref{subsec: 3.3}.
The best fitting model is the one that minimizes the reduced $\chi^2$, considering the composite flux described in the following equation \citep{2005AJ....130.1127F}:
\begin{equation}
f_{\lambda}=f_{\lambda,0}10^{-0.4 E(B-V)[k(\lambda-V)+R(V)]},
\label{eq: eq. 9}
\end{equation}

\noindent
where  $E(B-V)$ is the color excess, $k(\lambda-V)\equiv E(\lambda-V)/E(B-V)$ is the normalized extinction curve, $R(V)\equiv A(\lambda)/ E(B-V)$ is the ratio of total extinction to reddening at $V$ and $f_{\lambda,0}$ corresponds to the sum of intrinsic surface fluxes of both stars at the wavelength $\lambda$:
\begin{equation}
f_{\lambda,0}= \left(\frac{R_\textrm{c}}{d}\right)^{2} \left[ \left(\frac{R_\textrm{h}}{R_\textrm{c}}\right)^{2}  f_{\textrm{h},\lambda}+f_{\textrm{c},\lambda}\right],
\label{eq: eq. 10}
\end{equation}

\noindent
where the parameters $f_{\rm{h},\lambda}$ and $f_{\rm{c},\lambda}$ represent the fluxes of the hotter and colder star respectively, and the normalized extinction curve was calculated using  the prescriptions described in  \citet{1990ApJ...357..113M} and \citet{1989IAUS..135P...5C}.
The best fit resulted in a distance $d= 2300 \pm 50 \,\mathrm{pc}$ with a $\chi^2= 0.462$, close to the GAIA value mentioned in Section\,\ref{Sec: Sec. 1}, namely 2632 [+309 -251]  pc (Fig. \ref{fig:Fig. 9}). 


\begin{table*}
	\caption{Fluxes extracted from the Vizier Catalogue using a search radius of 2" around the position of DD\,CMa.}
	\begin{center}
		\resizebox{1.0\textwidth}{4cm}{
			\begin{tabular}{c c c l c c c l}
				\hline
				\noalign{\smallskip}
				\textrm {wavelength}   	& \textrm{Flux} 		& \textrm{Error}& \textrm{Filter ID}  & \textrm {wavelength}   	& \textrm{Flux} 		& \textrm{Error}& \textrm{Filter ID} \\
				\textrm{(\AA)}   		& \textrm{(erg s$^{-1}$ cm$^{-2}$ \AA$^{-1}$)}	&\textrm{(erg s$^{-1}$ cm$^{-2}$ \AA$^{-1}$)}        & & \textrm{(\AA)}   		&\textrm{(erg s$^{-1}$ cm$^{-2}$ \AA$^{-1}$)}	& \textrm{(erg s$^{-1}$ cm$^{-2}$ \AA$^{-1}$)}\\
				\hline
				\hline		
				\textrm{4203} 	& \textrm{1.40e-13}  &  \textrm{1.00e-14}  & \textrm{HIP:BT} &                                               \textrm{9091} 	& \textrm{1.45e-14}  &  \textrm{1.81e-16}  & \textrm{SkyMapper:z} \\
				\textrm{4442} 	& \textrm{1.35e-13}  &  \textrm{1.00e-14}  & \textrm{Johnson:B} &\textrm{9091} 	& \textrm{1.41e-14}  &  \textrm{1.45e-16}  & \textrm{SkyMapper:z} \\
				\textrm{4442} 	& \textrm{1.35e-13}  &  \textrm{1.32e-14}  & \textrm{Johnson:B} &\textrm{12390} 	& \textrm{1.17e-14}  &  \textrm{2.34e-16}  & \textrm{2MASS:J} \\
				\textrm{4442} 	& \textrm{1.12e-13}  &  \textrm{6.23e-15}  & \textrm{Johnson:B} &\textrm{12500} 	& \textrm{1.15e-14}  &  \textrm{1.34e-16}  & \textrm{Johnson:J} \\
				\textrm{4820} 	& \textrm{1.18e-13}  &  \textrm{4.52e-15}  & \textrm{SDSS:g'} &\textrm{12500} 	& \textrm{1.16e-14}  &  \textrm{2.11e-16}  & \textrm{Johnson:J} \\
				\textrm{4968} 	& \textrm{1.18e-13}  &  \textrm{1.21e-15}  & \textrm{SkyMapper:g} &\textrm{12500} 	& \textrm{1.17e-14}  &  \textrm{2.30e-16}  & \textrm{Johnson:J} \\
				\textrm{5050} 	& \textrm{9.64e-14}  &  \textrm{2.47e-15}  & \textrm{GAIA/GAIA2:Gbp} &\textrm{16300} 	& \textrm{4.48e-15}  &  \textrm{1.35e-16}  & \textrm{Johnson:H} \\
				\textrm{5319} 	& \textrm{9.39e-14}  &  \textrm{9.54e-15}  & \textrm{HIP:VT} &\textrm{16300} 	& \textrm{4.50e-15}  &  \textrm{1.13e-16}  & \textrm{Johnson:H} \\
				\textrm{5537} 	& \textrm{9.00e-14}  &  \textrm{6.26e-15}  & \textrm{Johnson:V} &\textrm{16500} 	& \textrm{4.44e-15}  &  \textrm{1.10e-16}  & \textrm{2MASS:H } \\
				\textrm{6040} 	& \textrm{6.70e-14}  &  \textrm{1.40e-15}  & \textrm{SkyMapper:r} &\textrm{21640} 	& \textrm{1.70e-15}  &  \textrm{3.20e-17}  & \textrm{2MASS:Ks} \\
				\textrm{6040} 	& \textrm{6.64e-14}  &  \textrm{1.15e-15}  & \textrm{SkyMapper:r} &\textrm{21900} 	& \textrm{1.59e-15}  &  \textrm{2.50e-17}  & \textrm{Johnson:K} \\
				\textrm{6230} 	& \textrm{5.91e-14}  &  \textrm{6.96e-16}  & \textrm{GAIA/GAIA2:G} &\textrm{33500} 	& \textrm{3.13e-16}  &  \textrm{5.34e-18}  & \textrm{WISE:W1} \\
				\textrm{6247} 	& \textrm{6.93e-14}  &  \textrm{9.22e-16}  & \textrm{SDSS:r'} &\textrm{33500} 	& \textrm{3.15e-16}  &  \textrm{8.01e-18}  & \textrm{WISE:W1} \\
				\textrm{7635} 	& \textrm{4.20e-14}  &  \textrm{1.18e-15}  & \textrm{SDSS:i'} &\textrm{46000} 	& \textrm{9.02e-17}  &  \textrm{1.84e-18}  & \textrm{WISE:W2} \\
				\textrm{7713} 	& \textrm{2.73e-14}  &  \textrm{3.53e-16}  & \textrm{SkyMapper:i} &\textrm{46000} 	& \textrm{9.17e-17}  &  \textrm{1.84e-18}  & \textrm{WISE:W2} \\
				\textrm{7713} 	& \textrm{2.76e-14}  &  \textrm{4.54e-16}  & \textrm{SkyMapper:i} &\textrm{11560} 	& \textrm{2.60e-18}  &  \textrm{2.92e-19}  & \textrm{WISE:W3} \\
				\textrm{7730} 	& \textrm{3.86e-14}  &  \textrm{9.55e-16}  & \textrm{GAIA/GAIA2:Grp} &\textrm{11560} 	& \textrm{3.25e-18}  &  \textrm{2.69e-19}  & \textrm{WISE:W3} \\	
				\hline
			\end{tabular}
		}
	\end{center}
	\label{Tab: Tab. 6}
\end{table*}

\section{Residual emission in Balmer lines}
\label{Sec: Sec. 6}

\noindent
We have subtracted the theoretical contribution of the hotter and cooler stars from the average spectrum. While the parameters used for the hotter star are those calculated in Section \ref{subsec: 3.4}, those for the secondary star arise from the light curve model. 
We model the cooler star with an ATLAS9 atmospheric model with effective temperature $T_\mathrm{c}= 11000 ~\mathrm{K}$, surface gravity $\log{g}= 3.5 ~\mathrm{dex}$, macro-turbulent velocity $v_\mathrm{mac}= 0.0 ~\mathrm{km\,s^{-1}}$ according to \citet{2004IAUS..224..131S}, a fixed micro-turbulent velocity $v_\mathrm{mic}=0.0 ~\mathrm{km\,s^{-1}}$, $v\sin{i}=  93.2 \pm 3.4 ~\mathrm{km\,s^{-1}}$,  a mixing length parameter of strong convection $l/H$= 1.25 and a fixed stellar metallicity $M/H$= 0.0. 

\begin{figure}
	\includegraphics[width=0.45\textwidth]{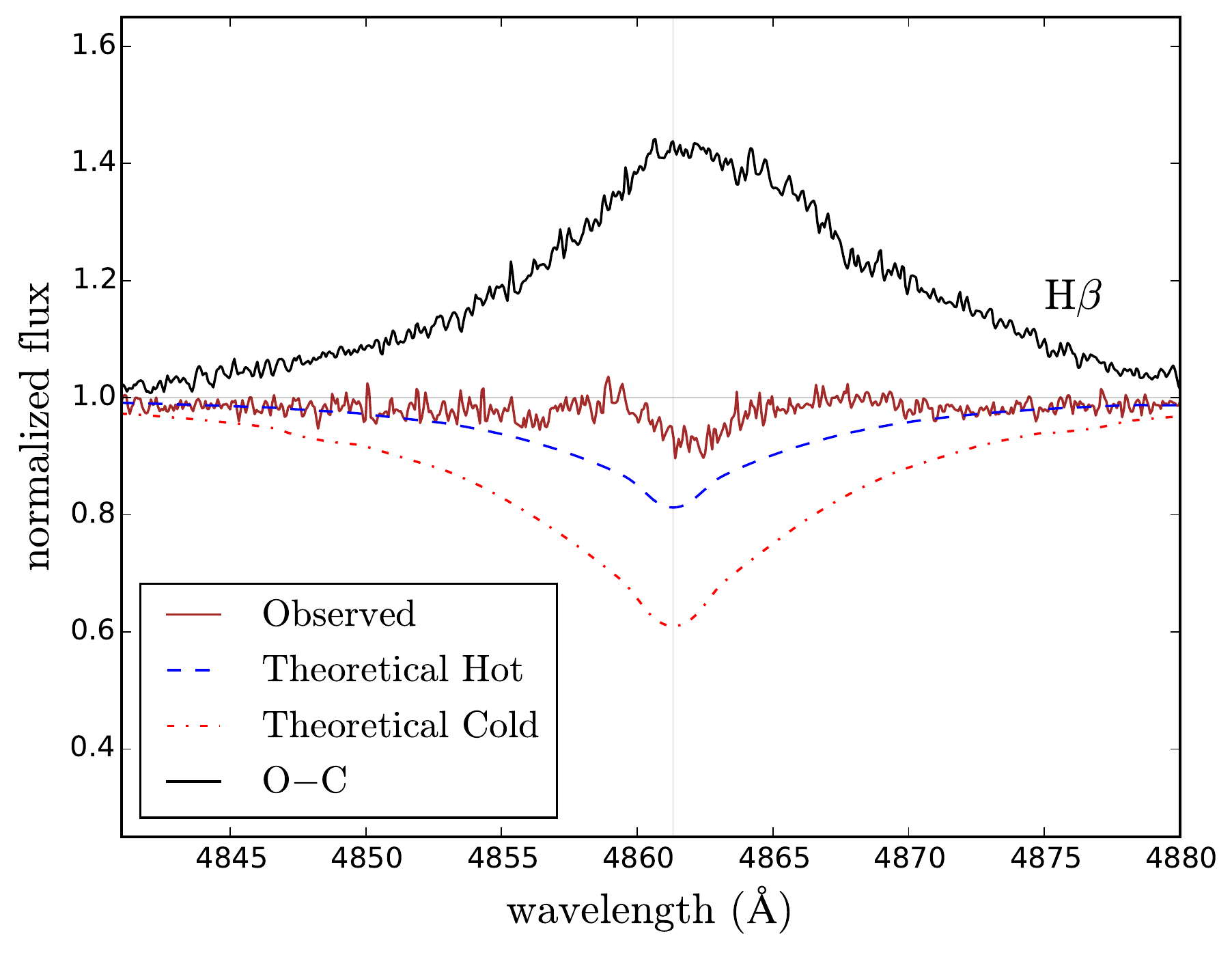}
	\includegraphics[width=0.45\textwidth]{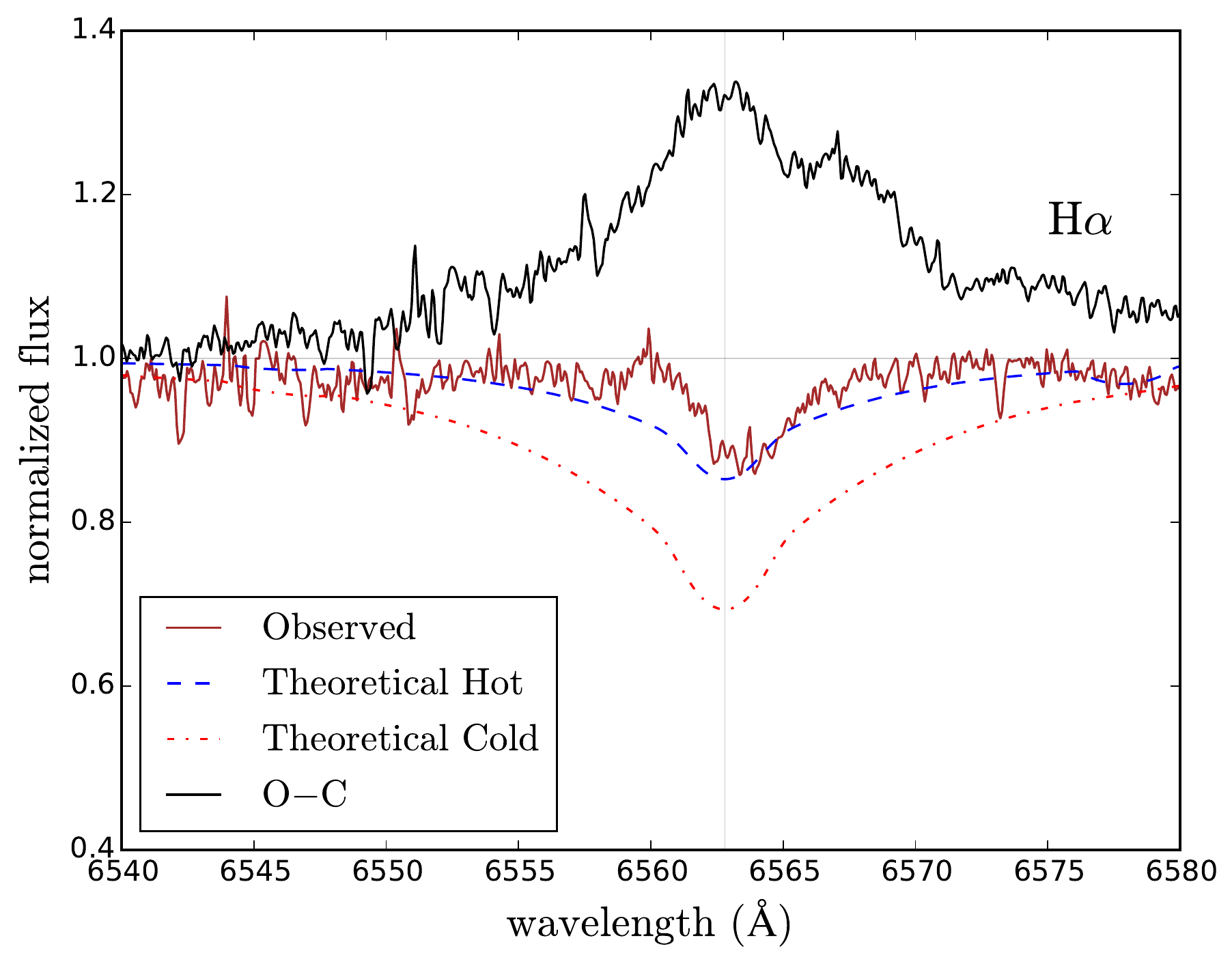}
	\caption{ Decomposed H$\beta$ and H$\alpha$ profiles of the observed spectrum using theoretical stellar components, at the orbital phase $\phi_{\mathrm{o}}= 0.775$. The brown line corresponds to the observed spectrum without  decomposition. The black line corresponds to the observed spectrum after subtracting the contributions of the theoretical  primary and secondary components, centered in H$\beta$ and H$\alpha$. These residual spectra were displaced vertically to fit the continuum.  The dashed blue line corresponds to the theoretical spectrum for a hot component. The dash-dotted red line corresponds to a theoretical secondary component. Both theoretical models were displaced to the right velocity and weighted by their relative contributions to the total flux at the orbital phase $\phi_{o}= 0.775$. 
	}
	\label{fig:Fig. 10}
\end{figure}

\begin{figure*}
	\begin{center}
		\includegraphics[trim=0.0cm 0.0cm 0.0cm 0.0cm,clip,width=0.5\textwidth,angle=0]{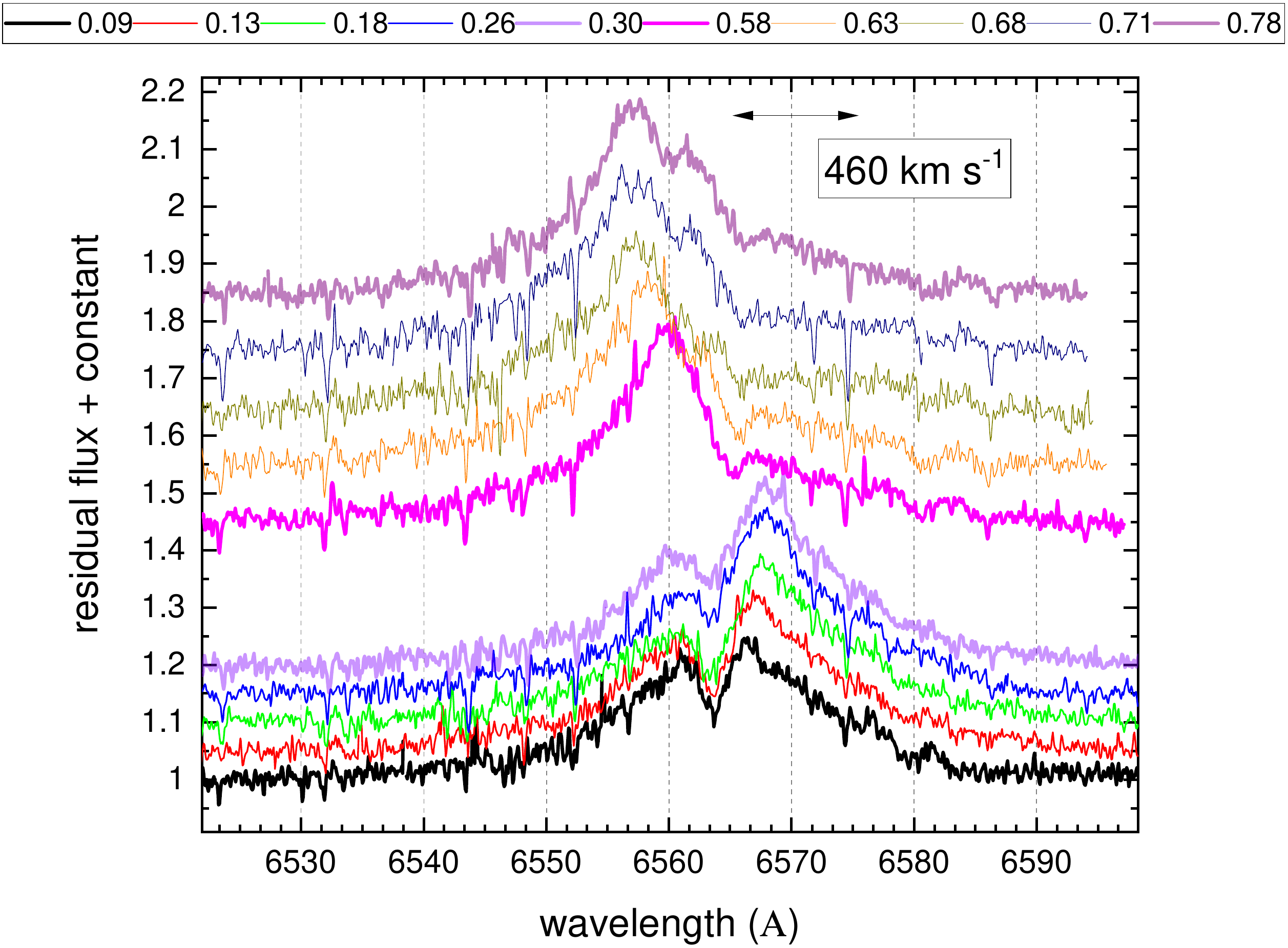}
		\includegraphics[trim=0.0cm 0.0cm 0.0cm 0.0cm,clip,width=0.47\textwidth,angle=0]{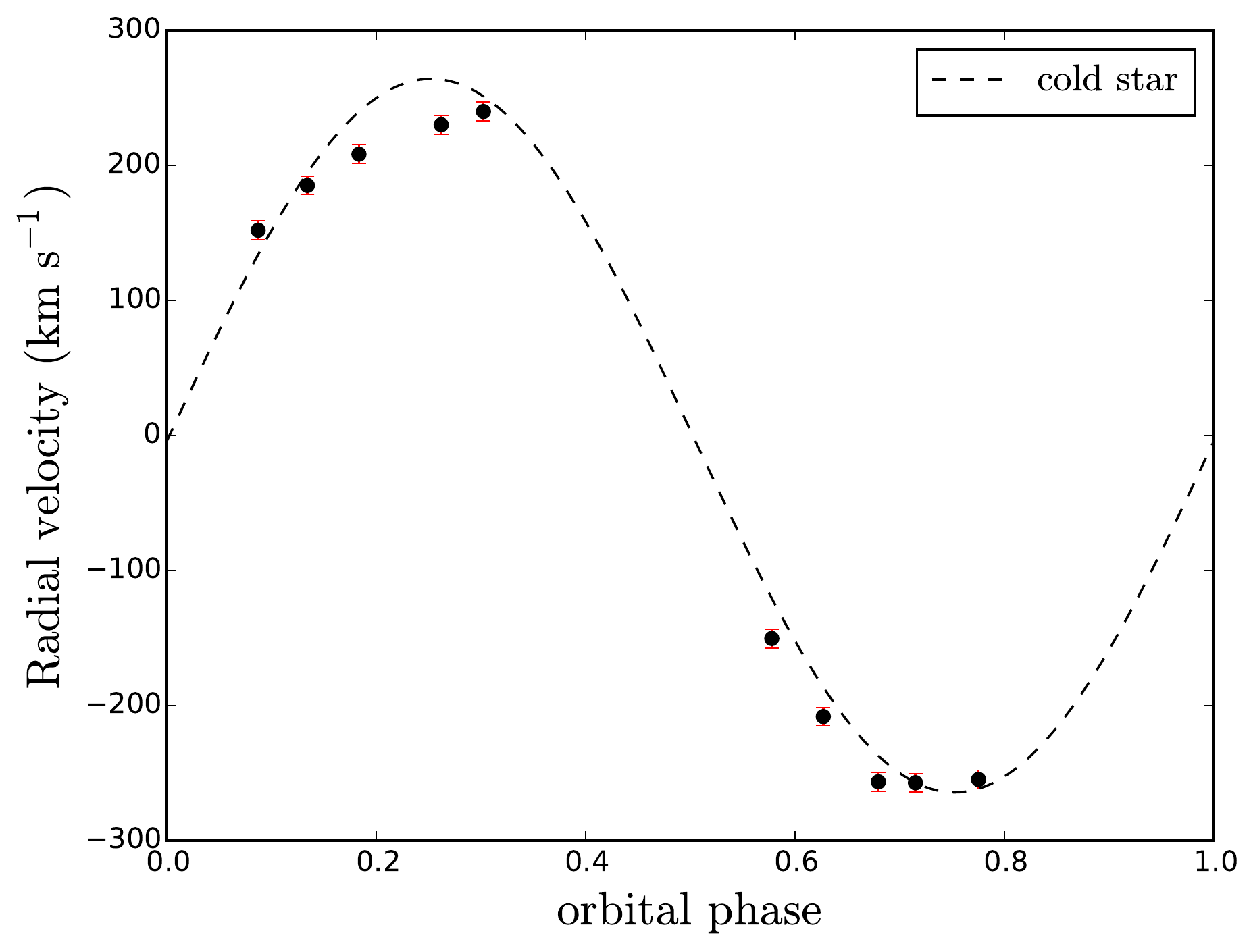}
	\end{center}
	\caption{Left: The H$\alpha$ residual emissions. The numbers in the upper scale indicate the orbital phases of the spectra. Right: Radial velocity of the H$\alpha$ residual emission peak and the best sinusoid fit, along with the donor theoretical radial velocity. }
	\label{fig:Fig. 11}
\end{figure*}

The spectra resulting  of the  subtraction of the stellar theoretical spectra reveal
broad emission peaks in H$\alpha$ and H$\beta$  (Fig. \ref{fig:Fig. 10}). A study of the
H$\alpha$ residual emission through the orbital cycle shows a complex and variable 
structure; a double, sometimes single, usually asymmetrical emission whose maximum intensity 
closely follows the velocity of the donor star (Fig. \ref{fig:Fig. 11}).

\ 

\section{Discussion}
\label{Sec: Sec. 7}

\begin{figure}
	\begin{center}
		\includegraphics[trim=0.2cm 0cm 0cm 0cm,clip,width=0.45\textwidth,angle=0]{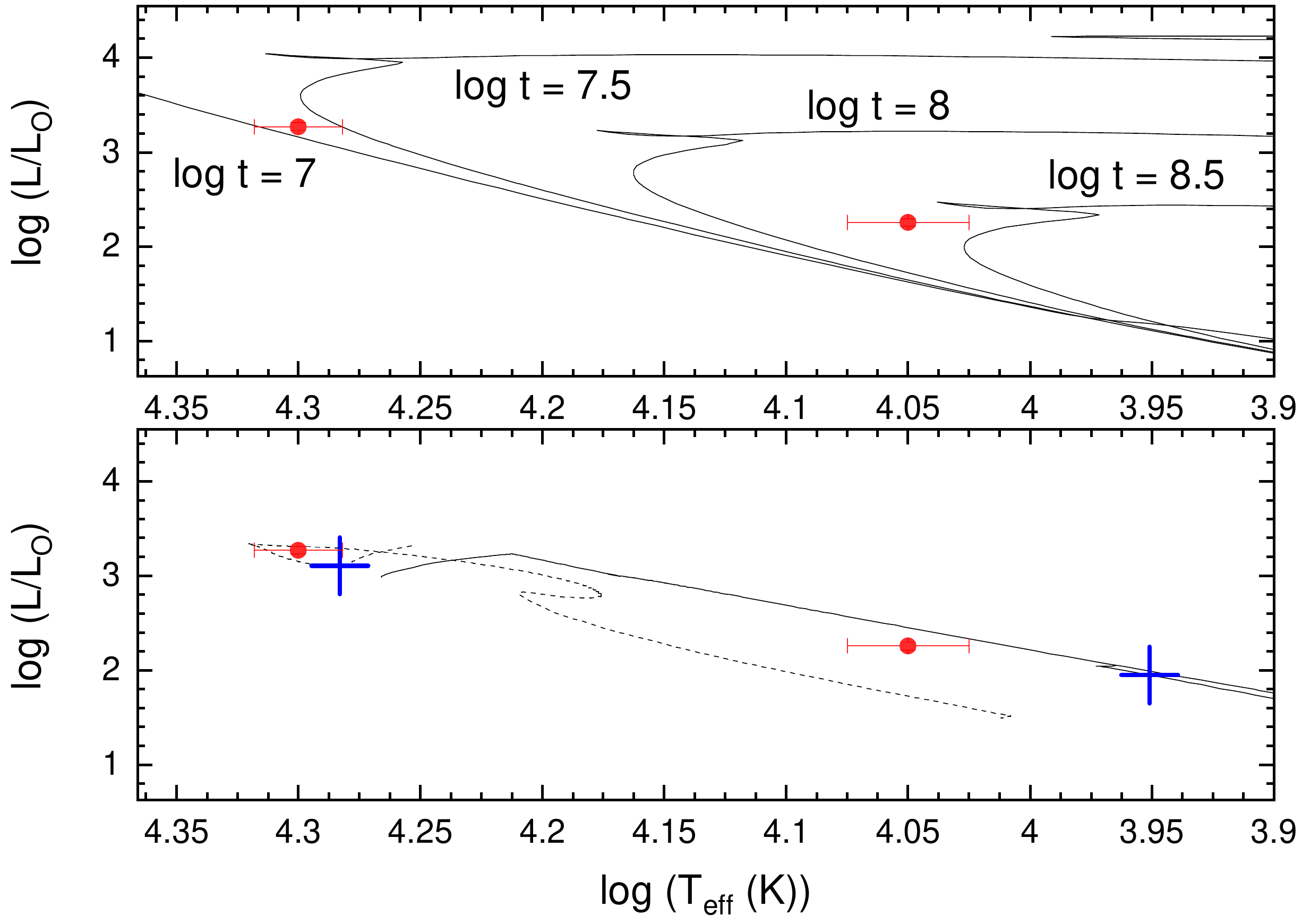}
	\end{center}
	\caption{The fractional radius, relative to the orbital separation, for primaries and disks in Algols with and without disks. Red and blue circles show primaries of Double Periodic Variables and W\,Serpentids, respectively. Triangles with the same color mark indicate disk radii. Symbol size for stellar radius is proportional to the system total mass. Below the circularization radius shown by the solid black line, a disk should be formed and below the dash-point a disk might be formed. The tidal radius indicates the maximum possible disk extension (upper dashed line). Semidetached Algol primaries from \citet{2010MNRAS.406.1071D} are also shown as black points. Adapted from \citet{2016MNRAS.455.1728M}.}
	\label{fig:Fig. 12}
\end{figure}

\noindent
The temperature of the hotter star, radius, and mass are consistent with the typical properties of main sequence stars, while the cold  star is clearly evolved showing a larger radius and smaller mass than expected for main sequences stars of the same temperature \citep{1988BAICz..39..329H}. 
We determine the stellar spectral types B\,9 and B\,2.5 following the calibration of \citet{1988BAICz..39..329H}. Therefore the system consists of an early B-type dwarf and a slight evolved 
late B-type giant. We notice that the fractional radius of the hotter star $R_{\mathrm{h}}/a = 0.24 \pm 0.01$ and the mass ratio $q = 0.268$ place the system into the area of \textquotedblleft direct impact\textquotedblright \,  semidetached binaries (Fig. \ref{fig:Fig. 12}). This means that if mass transfer occurs due to Roche-lobe overflow, the gas stream hits directly the more massive star. 

The line emission shoulders  observed in Balmer lines and the clear emissions observed after removing the theoretical underlying spectra, along with the infrared color excess, indicate the presence of circumstellar matter. The complex emission profiles observed in H$\alpha$ and H$\beta$ residual spectra do not correspond to the typical double-peaked emission of an optically thin accretion disk. We notice that the broad emission profiles have high velocities in their wings, probably indicating an origin in a region of rapidly rotating material. The Half Width at Half Maximum of the H$\alpha$ emission is $\sim$ 700 km\,s$^{-1}$.  This velocity is compatible with the critical velocity of early B-type stars \citep{2008A&A...478..467E}. It is in principle possible that the impact of the stream in the stellar surface speed it up to near critical velocity \citep{1981A&A...102...17P}.  However, 
the $v\sin\,i$ value determined from the fit to the helium lines - 80 km\,s$^{-1}$ - suggests that the hotter star rotates almost 
synchronously with the orbit; the expected projected synchronous velocity is 92 km\,s$^{-1}$. We notice that
other broadening mechanism besides rotation could be acting in the Balmer lines. 
For instance, in Be star disks, Thompson scattering of Balmer line photons by free envelope electrons produces extended H$\alpha$ emission wings \citep[][and references therein]{1990A&A...230..380D}. 

Quite interesting is the behavior of the velocity of the H$\alpha$ emission peak, that roughly follows the donor path.
We conjecture that material escaping from the donor through the inner Lagrangian point 
in the form of a gas stream is responsible for part of this emission. The fact that 
part of the material is observed receding and other part approaching the observer at almost all orbital phases covered by our spectra, reveal the complex structure of the gaseous envelope. This view is consistent with the semidetached nature of the binary and its direct-impact condition; it is also possible that part of the emission comes from the impact point of the stream on the gainer.  

We propose that DD\,CMa is a semidetached direct impact binary consisting of two B-type stars. The cooler star overfills its Roche lobe and transfers matter onto the hotter star though an accretion gas stream that hits directly the star's surface. 
The observed emission  in H$\alpha$ and H$\beta$ are probably in a gas stream between both stars, reflecting the process of mass exchange between the stars. 
The circumstellar matter  is also revealed by the infrared color excess found in the WISE photometry.

The error in the orbital period $\epsilon$ = 0\fd0000006, given by  Eq. \ref{eq: eq. 2}, might be interpreted as a possible drift of the orbital period between 
P$_o$ - $\epsilon$ and P$_o$ + $\epsilon$ in $\Delta$t = 18 years. This implies a possible change in the orbital period of less than roughly 2$\epsilon_P$/$\Delta$t = 6.6 $\times$ 10$^{-8}$ d/yr. 
For conservative mass transfer, this imposes an upper limit for the mass transfer rate. The expected period change in the conservative case is \citep{1963ApJ...138..471H}:
\begin{eqnarray}
\rm \frac{\dot{P}_o}{P_o} = -3 \dot{\cal M_{\rm_c}} \left( \frac{1}{\cal M_{\rm_{c}}}-\frac{1}{\cal M_{\rm_{h}}} \right).
\label{eq: eq. 11}
\end{eqnarray}

\noindent
Using the aforementioned numbers and the derived stellar masses, we determine $\dot{\cal M_{\rm_c}}$ $<$ 2.55 $\times$ 10$^{-8}$  $\rm M_{\odot}$/yr.  We have not found evidence for a change in the orbital period, but in case it does change, the upper limit we have found imposes a restriction for the mass transfer rate, in the conservative case, reflected in the above figure.

\begin{figure}
	\begin{center}
		\includegraphics[trim=0.2cm 0cm 0cm 0cm,clip,width=0.45\textwidth,angle=0]{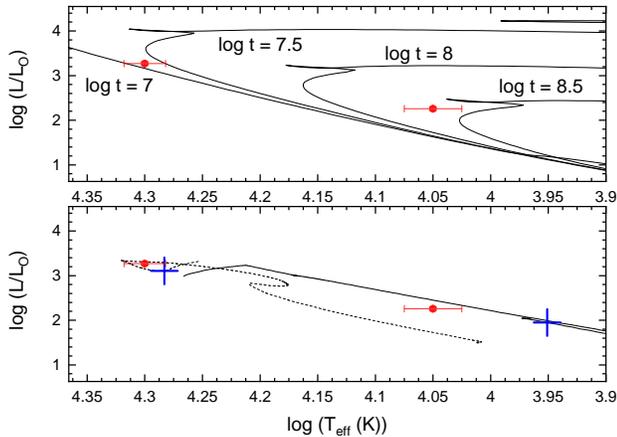}
	\end{center}
	\caption{Top panel:  PARSEC isochrones for different ages compared with the observed parameters for the binary stellar components. Bottom panel:
		The evolutionary tracks for an initially 6+2.4 M$_{\odot}$ binary according to  \citet{2016A&A...592A.151V} compared with the component luminosities and temperatures.  Crosses show the position of the model that best matches the observed stellar luminosities, temperatures, masses and radii and the orbital period.}
	\label{fig:Fig. 13}
\end{figure}

When comparing the stellar luminosities and temperatures with those predicted by PARSEC\footnote{http://stev.oapd.inaf.it/cgi-bin/cmd} isochrones for solar-composition stellar models \citep{2012MNRAS.427..127B}, we find that the hotter star is located near the main sequence whereas the cooler star is slightly evolved. 
Both stars cannot fit the same age simultaneously, when considering single star evolution models
(Fig. \ref{fig:Fig. 13}; top panel).  However, the position of the stars in the luminosity-temperature diagram can be understood in terms of mass exchange in a close interacting binary. We searched for the best binary model among those described in  \citet{2016A&A...592A.151V} following the method described in \citet{2013MNRAS.432..799M}, which consists in a $\chi^2$ minimization of modeled and observed luminosities, masses, temperatures, radii and periods.

The best fit is obtained with a 2\fd026 orbital period binary of initial masses
6 and 2.4 M$_{\odot}$ and initial orbital period 1\fd5 days, whose less massive star is filling its Roche lobe, 
igniting hydrogen in its core. The binary has an age of 50 Myr and the cooler star is currently transferring matter onto the more massive star at a rate 
$\dot{\cal M_{\rm_c}}$ = 1.1 $\times$ 10$^{-6}$  $\rm M_{\odot}$/yr (Fig. \ref{fig:Fig. 13}; lower panel). 
As the grid of models is rather coarse - not including all posible initial parameters - and the model is simplified, since it does not consider the effect of circumstellar matter on the stellar luminosities and temperatures, we cannot expect a perfect match. In this sense our result is qualitative rather quantitative.  This explains  the  discrepancy among the mass transfer rates, for instance.  We include the best fit evolutionary track just to show that the current stellar and orbital parameters can in principle be reproduced by binary evolution considering mass exchange between the stellar components. In this view, the
initially less massive star evolved to higher masses and luminosities following a track almost  parallel to the main sequence, while acquiring mass from the initially more massive star.
The initially more massive star followed a completely different track, decaying in mass, temperature and luminosity. We observe that the above changes might explain the position of the binary components in the luminosity-temperature diagram, corroborating our general picture of mass exchange described in previous paragraphs.

\section{Conclusions}
\label{Sec: Sec. 8}
In this work we have presented a detailed spectroscopic and photometric study of the eclipsing binary DD\,CMa.
This is the first spectroscopic study of this system and our main results are the following:

\begin{itemize}
	
	\item We analyzed main eclipse times spanning 84 years, and find an improved orbital period of 2\fd0084530(6).

	\item From the analysis of the spectral energy distribution, we find a distance of
	2300 pc with a formal error of 50 pc, close to the reported value based on the GAIA DR2, viz. 2632  [+309 -251] pc. 
	
	\item  From the analysis of the radial velocities we have found a circular orbit, $q$= 0.268 $\pm$ 0.01, $K_\mathrm{h} = 70.6 \pm 1.2 ~\mathrm{km\,s^{-1}}$ and $\gamma =  0.4 \pm 0.6 ~\mathrm{km\,s^{-1}}$. In addition, its companion shown a semi-amplitude $K_\mathrm{c}=264.2 \pm 5.4 ~\mathrm{km\,s^{-1}}$. 
	
	\item We find H$\alpha$ absorption flanked by weak emission shoulders. This finding, along with the clear detection of Balmer emission in residual spectra and the infrared excess observed in WISE photometry, suggest the presence of  circumstellar emitting gas.  The radial velocities of the residual emission suggest a partial origin in the gas stream. The above suggests that DD\,CMa is in a mass exchange evolutionary stage.
	
	\item We solved the inverse problem to derive the parameters producing the best match between a theoretical light curve, and the observed one. The best model of the $V$-band ASAS-SN light curve includes a slightly evolved late B-type star of $M_{\mathrm{c}} = 1.7 \pm 0.1 ~\mathrm{M_{\odot}}$, $T_{\mathrm{c}} = 11350 \pm 100 ~\mathrm{K}$, and $R_{\mathrm{c}} = 3.7 \pm 0.1 ~\mathrm{R_{\odot}}$. It also includes an early B-type star with $M_\mathrm{h} = 6.4 \pm 0.1 ~\mathrm{M_{\odot}}$, $T_\mathrm{h} = 20000 \pm 500 ~\mathrm{K}$, and $R_\mathrm{h} = 3.2 \pm 0.1 ~\mathrm{R_\mathrm{\odot}}$. The inclination of the orbit is 81\fdg8.
	
	\item When comparing the obtained orbital and stellar parameters with those of published evolutionary tracks for intermediate mass binaries, we find
	that they can be reproduced considering the mass exchange occurring when the initially more massive star fills its Roche lobe as result of its nuclear evolution.

\end{itemize}


\section{Acknowledgments}

\noindent
We thank the anonymous referee who provided useful insights on the first version of this paper. J.R., REM and DS acknowledge support by Fondecyt 1190621, Fondecyt 1201280 and the BASAL (CATA) PFB-06/2007. GD acknowledges the financial support of the Ministry of Education and Science of the Republic of Serbia through the project 176004 \textquotedblleft Stellar  Physics \textquotedblright and the financial support of the  Ministry of Education, Science and Technological Development of the Republic of Serbia through the contract N$^{\rm{o}}$ 451-03-68/2020-14/200002. M. Cabezas acknowledges support by Astronomical Institute of the Czech Academy of Sciences through the project RVO 67985815.  IA and M. Cur\'e acknowledges support from Fondecyt 1190485. IA is also grateful for the support from Fondecyt Iniciaci\'on 11190147.

\bibliography{b63}{}

\begin{thebibliography}{}
\expandafter\ifx\csname natexlab\endcsname\relax\def\natexlab#1{#1}\fi

\bibitem[{{Bailer-Jones} {et~al.}(2018){Bailer-Jones}, {Rybizki}, {Fouesneau},
  {Mantelet}, \& {Andrae}}]{2018AJ....156...58B}
{Bailer-Jones}, C.~A.~L., {Rybizki}, J., {Fouesneau}, M., {Mantelet}, G., \&
  {Andrae}, R. 2018, \aj, 156, 58

\bibitem[{{Bayo} {et~al.}(2008){Bayo}, {Rodrigo}, {Barrado Y Navascu{\'e}s},
  {Solano}, {Guti{\'e}rrez}, {Morales-Calder{\'o}n}, \&
  {Allard}}]{2008A&A...492..277B}
{Bayo}, A., {Rodrigo}, C., {Barrado Y Navascu{\'e}s}, D., {et~al.} 2008, \aap,
  492, 277

\bibitem[{{Bressan} {et~al.}(2012){Bressan}, {Marigo}, {Girardi}, {Salasnich},
  {Dal Cero}, {Rubele}, \& {Nanni}}]{2012MNRAS.427..127B}
{Bressan}, A., {Marigo}, P., {Girardi}, L., {et~al.} 2012, \mnras, 427, 127

\bibitem[{{Cardelli} {et~al.}(1989){Cardelli}, {Clayton}, \&
  {Mathis}}]{1989IAUS..135P...5C}
{Cardelli}, J.~A., {Clayton}, G.~C., \& {Mathis}, J.~S. 1989, in Interstellar
  Dust, ed. L.~J. {Allamandola} \& A.~G.~G.~M. {Tielens}, Vol. 135, 5--10

\bibitem[{{Castelli} \& {Kurucz}(2003)}]{2003IAUS..210P.A20C}
{Castelli}, F., \& {Kurucz}, R.~L. 2003, in Modelling of Stellar Atmospheres,
  ed. N.~{Piskunov}, W.~W. {Weiss}, \& D.~F. {Gray}, Vol. 210, A20

\bibitem[{{Charbonneau}(1995)}]{1995ApJS..101..309C}
{Charbonneau}, P. 1995, \apjs, 101, 309

\bibitem[{{Dachs} \& {Rohe}(1990)}]{1990A&A...230..380D}
{Dachs}, J., \& {Rohe}, C. 1990, \aap, 230, 380

\bibitem[{{de Mink} {et~al.}(2014){de Mink}, {Sana}, {Langer}, {Izzard}, \&
  {Schneider}}]{2014ApJ...782....7D}
{de Mink}, S.~E., {Sana}, H., {Langer}, N., {Izzard}, R.~G., \& {Schneider},
  F.~R.~N. 2014, \apj, 782, 7

\bibitem[{{Dervi{\textcommabelow s}o{\v{g}}lu}
  {et~al.}(2010){Dervi{\textcommabelow s}o{\v{g}}lu}, {Tout}, \&
  {Ibano{\v{g}}lu}}]{2010MNRAS.406.1071D}
{Dervi{\textcommabelow s}o{\v{g}}lu}, A., {Tout}, C.~A., \& {Ibano{\v{g}}lu},
  C. 2010, \mnras, 406, 1071

\bibitem[{{Deurinck}(1949)}]{1949PLAGL..12E..17D}
{Deurinck}, R. 1949, Publications du Laboratoire d'Astronomie et de Geodesie de
  l'Universite de Louvain, 12, E17

\bibitem[{{Djurasevic}(1992)}]{1992Ap&SS.197...17D}
{Djurasevic}, G. 1992, \apss, 197, 17

\bibitem[{{Djura{\v{s}}evi{\'c}}(1996)}]{1996Ap&SS.243..413D}
{Djura{\v{s}}evi{\'c}}, G. 1996, \apss, 243, 413

\bibitem[{{Djura{\v{s}}evi{\'c}} {et~al.}(2013){Djura{\v{s}}evi{\'c}},
  {Ba{\textcommabelow s}t{\"u}rk}, {Latkovi{\'c}}, {Y{\i}lmaz},
  {{\c{C}}al{\i}{\textcommabelow s}kan}, {Tanr{\i}verdi}, {{\c{S}}enavc{\i}},
  {K{\i}l{\i}{\c{c}}o{\v{g}}lu}, \& {Ekmek{\c{c}}i}}]{2013AJ....145...80D}
{Djura{\v{s}}evi{\'c}}, G., {Ba{\textcommabelow s}t{\"u}rk}, {\"O}.,
  {Latkovi{\'c}}, O., {et~al.} 2013, \aj, 145, 80

\bibitem[{{Eggleton}(2006)}]{2006epbm.book.....E}
{Eggleton}, P. 2006, {Evolutionary Processes in Binary and Multiple Stars}

\bibitem[{{Eggleton}(1983)}]{1983ApJ...268..368E}
{Eggleton}, P.~P. 1983, \apj, 268, 368

\bibitem[{{Ekstr{\"o}m} {et~al.}(2008){Ekstr{\"o}m}, {Meynet}, {Maeder}, \&
  {Barblan}}]{2008A&A...478..467E}
{Ekstr{\"o}m}, S., {Meynet}, G., {Maeder}, A., \& {Barblan}, F. 2008, \aap,
  478, 467

\bibitem[{{Fitzpatrick} \& {Massa}(2005)}]{2005AJ....130.1127F}
{Fitzpatrick}, E.~L., \& {Massa}, D. 2005, \aj, 130, 1127

\bibitem[{{Gonz{\'a}lez} \& {Levato}(2006)}]{2006A&A...448..283G}
{Gonz{\'a}lez}, J.~F., \& {Levato}, H. 2006, \aap, 448, 283

\bibitem[{{Harmanec}(1988)}]{1988BAICz..39..329H}
{Harmanec}, P. 1988, Bulletin of the Astronomical Institutes of Czechoslovakia,
  39, 329

\bibitem[{{Herbig}(1993)}]{1993ApJ...407..142H}
{Herbig}, G.~H. 1993, \apj, 407, 142

\bibitem[{{Hilditch}(2001)}]{2001icbs.book.....H}
{Hilditch}, R.~W. 2001, {An Introduction to Close Binary Stars}

\bibitem[{{Huang}(1963)}]{1963ApJ...138..471H}
{Huang}, S.-S. 1963, \apj, 138, 471

\bibitem[{{Hubber} \& {Whitworth}(2005)}]{2005A&A...437..113H}
{Hubber}, D.~A., \& {Whitworth}, A.~P. 2005, \aap, 437, 113

\bibitem[{{Jayasinghe} {et~al.}(2019){Jayasinghe}, {Stanek}, {Kochanek},
  {Shappee}, {Holoien}, {Thompson}, {Prieto}, {Dong}, {Pawlak}, {Pejcha},
  {Shields}, {Pojmanski}, {Otero}, {Britt}, \& {Will}}]{2019MNRAS.486.1907J}
{Jayasinghe}, T., {Stanek}, K.~Z., {Kochanek}, C.~S., {et~al.} 2019, \mnras,
  486, 1907

\bibitem[{{Kochanek} {et~al.}(2017){Kochanek}, {Shappee}, {Stanek}, {Holoien},
  {Thompson}, {Prieto}, {Dong}, {Shields}, {Will}, {Britt}, {Perzanowski}, \&
  {Pojma{\'n}ski}}]{2017PASP..129j4502K}
{Kochanek}, C.~S., {Shappee}, B.~J., {Stanek}, K.~Z., {et~al.} 2017, \pasp,
  129, 104502

\bibitem[{{Lucy} \& {Sweeney}(1971)}]{1971AJ.....76..544L}
{Lucy}, L.~B., \& {Sweeney}, M.~A. 1971, \aj, 76, 544

\bibitem[{Marquardt(1963)}]{1963SIAM...11..431}
Marquardt, D.~W. 1963, Journal of the Society for Industrial and Applied
  Mathematics, 11, 431

\bibitem[{{Martin} \& {Whittet}(1990)}]{1990ApJ...357..113M}
{Martin}, P.~G., \& {Whittet}, D.~C.~B. 1990, \apj, 357, 113

\bibitem[{{Mennickent} \& {Djura{\v{s}}evi{\'c}}(2013)}]{2013MNRAS.432..799M}
{Mennickent}, R.~E., \& {Djura{\v{s}}evi{\'c}}, G. 2013, \mnras, 432, 799

\bibitem[{{Mennickent} {et~al.}(2012){Mennickent}, {Djura{\v{s}}evi{\'c}},
  {Ko{\l}aczkowski}, \& {Michalska}}]{2012MNRAS.421..862M}
{Mennickent}, R.~E., {Djura{\v{s}}evi{\'c}}, G., {Ko{\l}aczkowski}, Z., \&
  {Michalska}, G. 2012, \mnras, 421, 862

\bibitem[{{Mennickent} {et~al.}(2016){Mennickent}, {Otero}, \&
  {Ko{\l}aczkowski}}]{2016MNRAS.455.1728M}
{Mennickent}, R.~E., {Otero}, S., \& {Ko{\l}aczkowski}, Z. 2016, \mnras, 455,
  1728

\bibitem[{{Packet}(1981)}]{1981A&A...102...17P}
{Packet}, W. 1981, \aap, 102, 17

\bibitem[{{Perryman} {et~al.}(1997){Perryman}, {Lindegren}, {Kovalevsky},
  {Hog}, {Bastian}, {Bernacca}, {Creze}, {Donati}, {Grenon}, {Grewing}, {van
  Leeuwen}, {van der Marel}, {Mignard}, {Murray}, {Le Poole}, {Schrijver},
  {Turon}, {Arenou}, {Froeschle}, \& {Petersen}}]{1997A&A...323L..49P}
{Perryman}, M.~A.~C., {Lindegren}, L., {Kovalevsky}, J., {et~al.} 1997, \aap,
  500, 501

\bibitem[{{Pickering}(1890)}]{1890MNRAS..50..296P}
{Pickering}, E.~C. 1890, \mnras, 50, 296

\bibitem[{{Pojmanski}(1997)}]{1997AcA....47..467P}
{Pojmanski}, G. 1997, \actaa, 47, 467

\bibitem[{{Rosales G.} \& {Mennickent}(2017)}]{2017IBVS.6207....1R}
{Rosales G.}, J., \& {Mennickent}, R.~E. 2017, Information Bulletin on Variable
  Stars, 6207, 1

\bibitem[{{Sana} {et~al.}(2012){Sana}, {de Mink}, {de Koter}, {Langer},
  {Evans}, {Gieles}, {Gosset}, {Izzard}, {Le Bouquin}, \&
  {Schneider}}]{2012Sci...337..444S}
{Sana}, H., {de Mink}, S.~E., {de Koter}, A., {et~al.} 2012, Science, 337, 444

\bibitem[{{Shappee} {et~al.}(2014){Shappee}, {Prieto}, {Grupe}, {Kochanek},
  {Stanek}, {De Rosa}, {Mathur}, {Zu}, {Peterson}, {Pogge}, {Komossa}, {Im},
  {Jencson}, {Holoien}, {Basu}, {Beacom}, {Szczygie{\l}}, {Brimacombe},
  {Adams}, {Campillay}, {Choi}, {Contreras}, {Dietrich}, {Dubberley},
  {Elphick}, {Foale}, {Giustini}, {Gonzalez}, {Hawkins}, {Howell}, {Hsiao},
  {Koss}, {Leighly}, {Morrell}, {Mudd}, {Mullins}, {Nugent}, {Parrent},
  {Phillips}, {Pojmanski}, {Rosing}, {Ross}, {Sand}, {Terndrup}, {Valenti},
  {Walker}, \& {Yoon}}]{2014ApJ...788...48S}
{Shappee}, B.~J., {Prieto}, J.~L., {Grupe}, D., {et~al.} 2014, \apj, 788, 48

\bibitem[{{Skrutskie} {et~al.}(2006){Skrutskie}, {Cutri}, {Stiening},
  {Weinberg}, {Schneider}, {Carpenter}, {Beichman}, {Capps}, {Chester},
  {Elias}, {Huchra}, {Liebert}, {Lonsdale}, {Monet}, {Price}, {Seitzer},
  {Jarrett}, {Kirkpatrick}, {Gizis}, {Howard}, {Evans}, {Fowler}, {Fullmer},
  {Hurt}, {Light}, {Kopan}, {Marsh}, {McCallon}, {Tam}, {Van Dyk}, \&
  {Wheelock}}]{2006AJ....131.1163S}
{Skrutskie}, M.~F., {Cutri}, R.~M., {Stiening}, R., {et~al.} 2006, \aj, 131,
  1163

\bibitem[{{Smalley}(2004)}]{2004IAUS..224..131S}
{Smalley}, B. 2004, in The A-Star Puzzle, ed. J.~{Zverko}, J.~{Ziznovsky},
  S.~J. {Adelman}, \& W.~W. {Weiss}, Vol. 224, 131--138

\bibitem[{{Stellingwerf}(1978)}]{1978ApJ...224..953S}
{Stellingwerf}, R.~F. 1978, \apj, 224, 953

\bibitem[{{Sterken}(2005)}]{2005ASPC..335....3S}
{Sterken}, C. 2005, in Astronomical Society of the Pacific Conference Series,
  Vol. 335, The Light-Time Effect in Astrophysics: Causes and cures of the O-C
  diagram, ed. C.~{Sterken}, 3

\bibitem[{{Tody}(1993)}]{1993ASPC...52..173T}
{Tody}, D. 1993, in Astronomical Society of the Pacific Conference Series,
  Vol.~52, Astronomical Data Analysis Software and Systems II, ed. R.~J.
  {Hanisch}, R.~J.~V. {Brissenden}, \& J.~{Barnes}, 173

\bibitem[{{Ulrich} \& {Burger}(1976)}]{1976ApJ...206..509U}
{Ulrich}, R.~K., \& {Burger}, H.~L. 1976, \apj, 206, 509

\bibitem[{{Van Rensbergen} \& {De Greve}(2016)}]{2016A&A...592A.151V}
{Van Rensbergen}, W., \& {De Greve}, J.~P. 2016, \aap, 592, A151

\bibitem[{{Wilson} \& {Devinney}(1971)}]{1971ApJ...166..605W}
{Wilson}, R.~E., \& {Devinney}, E.~J. 1971, \apj, 166, 605

\bibitem[{{Wright} {et~al.}(2010){Wright}, {Eisenhardt}, {Mainzer}, {Ressler},
  {Cutri}, {Jarrett}, {Kirkpatrick}, {Padgett}, {McMillan}, {Skrutskie},
  {Stanford}, {Cohen}, {Walker}, {Mather}, {Leisawitz}, {Gautier}, {McLean},
  {Benford}, {Lonsdale}, {Blain}, {Mendez}, {Irace}, {Duval}, {Liu}, {Royer},
  {Heinrichsen}, {Howard}, {Shannon}, {Kendall}, {Walsh}, {Larsen}, {Cardon},
  {Schick}, {Schwalm}, {Abid}, {Fabinsky}, {Naes}, \&
  {Tsai}}]{2010AJ....140.1868W}
{Wright}, E.~L., {Eisenhardt}, P. R.~M., {Mainzer}, A.~K., {et~al.} 2010, \aj,
  140, 1868

\end{thebibliography}
\bibliographystyle{aasjournal}

\end{document}